\documentclass[journal]{IEEEtran}
\usepackage[dvips]{graphicx}
\usepackage{indentfirst}
\usepackage{amsmath}
\usepackage{amssymb}
\usepackage{threeparttable}
\usepackage{multirow}
\usepackage{subfigure}
\usepackage{xcolor}
\usepackage{cite}
\usepackage{enumerate}
\usepackage{algorithm}
\usepackage{algorithmic}
\usepackage{setspace}
\usepackage{multicol}
\usepackage{stfloats}
\usepackage{array}

\newtheorem{Definition}{Definition}
\newtheorem{Theorem}{Theorem}
\newtheorem{Remark}{Remark}

\newtheorem{Lemma}{Lemma}

\newtheorem{Example}{Example}

\begin{document}

\title{Reinforcement Learning Based Power Control for Reliable Mission-Critical Wireless Transmission}

\author{
Chongtao Guo,~\IEEEmembership{Member,~IEEE},
Zhengchao Li,
Le Liang,~\IEEEmembership{Member,~IEEE}, and
Geoffrey Ye Li,~\IEEEmembership{Fellow,~IEEE}

\thanks{Copyright (c) 20xx IEEE. Personal use of this material is permitted. However, permission to use this material for any other purposes must be obtained from the IEEE by sending a request to pubs-permissions@ieee.org.}

\thanks{This work was supported in part by Natural Science Foundation of Shenzhen under Grant JCYJ20190808114213987, in part by University Stability Support Program of Shenzhen under Grant 20200812112423002, in part by  Natural Science Foundation of Jiangsu Province under Grant BK20220810, and in part by Natural Science Foundation of China under Grant 62201145 and Grant 62231019. \emph{(Corresponding author: Le Liang.)}}

\thanks{
C. Guo is with the Guangdong Key Laboratory of Intelligent Information Processing, College of Electronics and Information Engineering, Shenzhen University, Shenzhen 518060, China (E-mail: ctguo@szu.edu.cn).

Z. Li is with Huawei Technologies Co., Ltd, Dongguan 523820, China (E-mail: zhengcli@qq.com).

L. Liang is with the National Mobile Communications Research Laboratory and Frontiers Science Center for Mobile Information Communication and Security, Southeast University, Nanjing 210096, and also with the Purple Mountain Laboratories, Nanjing 211111, China (e-mail: lliang@seu.edu.cn).

G. Y. Li is with the Dept. of Electrical and Electronic Engineering, Imperial College London, Exhibition Road, London SW7 2AZ, UK (E-mail: geoffrey.li@imperial.ac.uk).
}

}
\maketitle

\begin{abstract}
In this paper, we investigate sequential power allocation over fast varying channels for mission-critical applications, aiming to minimize the expected sum power while guaranteeing the transmission success probability.
In particular, a reinforcement learning framework is constructed with appropriate reward design so that the optimal policy maximizes the Lagrangian of the primal problem, where the maximizer of the Lagrangian is shown to have several good properties.
For the model-based case, a fast converging algorithm is proposed to find the optimal Lagrange multiplier and thus the corresponding optimal policy.
For the model-free case, we develop a three-stage strategy, composed in order of online sampling, offline learning, and online operation, where a backward Q-learning with full exploitation of sampled channel realizations is designed to accelerate the learning process.
According to our simulation, the proposed reinforcement learning framework can solve the primal optimization problem from the dual perspective.
Moreover, the model-free strategy achieves a performance close to that of the optimal model-based algorithm.
\end{abstract}


\section{ Introduction }\label{Section_Introduction}

\IEEEPARstart{D}{ata} packets in mission-critical applications, such as health monitoring, autonomous driving, haptic interaction, and
factory automation, usually contain vital information that demands strict performance bound on packet delivery reliability and latency.
In the example of automatic driving, surrounding information observed by self-equipped devices, such as camera and radar, is not enough for intelligent vehicles to make smart decisions.
Information, such as common awareness messages (CAM) and decentralized environment notification messages (DENM), needs to be delivered in the Internet of Vehicles (IoV) \cite{2017-MVT-CoopITSinEurope}, which helps vehicles  know nearby traffic lights, location of pedestrians, speed of neighboring vehicles, and emergency alert, etc.
Such information is often safety-critical since  failure or delay of packet transmission may cause catastrophic consequences, such as severe road accidents \cite{2022-Network-AoILatencyReliability}.
Another use case is a specific Internet of Things (IoT) called wireless sensor and actuator networks in the area of factory automation, where the actuator must accurately and quickly respond to the collected data \cite{2012-CST-SurveyMACProtocolMissionCriticalApp}.
For example, in chemical production process, sensors monitor pressure in pipes and send the pressure information to an actuator in control of a valve.
Definitely, data delivery from the sensor to actuator should be successfully done before a tight deadline to make the control loop well implemented and guarantee production safety.

In mission-critical tasks, packets are usually periodically generated at the transmitter and are required to be successfully delivered  to a receiver with a deadline before the next cycle.
Moreover, the packets that fail to reach the destination due to, e.g., deep fading of wireless channels, will become outdated thereafter and  will be dropped prior to the next cycle.
For instance, in the fifth generation (5G) enabled automatic driving system, each vehicle may need to periodically and timely report its  location and speed information to its neighbours, indicating that the packets will be dropped once they have been scheduled to transmit whether the transmission is successful or not.
It can be expected that the latency requirement here is most equivalent to the reliability requirement, where the reliability is generally defined as the probability that a given amount of data is successfully delivered from a source node to a sink node within a certain time period \cite{2018-Proc-URLLWirelessCommunTailRiskScale}.
In particular, the 3rd generation partnership project (3GPP) defines a general reliability requirement for the 5G ultra-reliable low-latency communication (URLLC) cases  such that $32$ bytes of data must be transmitted within $1$ ms with a success probability of $1-10^{-5}$ \cite{2017-3GPP-22261}.
In the following, we will  introduce the related works on reliability guarantee in wireless mission-critical systems and highlight the contributions of this work.

\subsection{Related Works}

It is usually not difficult to guarantee reliable mission-critical wireless data transmission by appropriate resource allocation when the channel remains constant during the data transmission period and accurate channel state information (CSI) is available.
For the power and rate allocation scheme proposed by \cite{2019-TCOM-JointPowerControlInSIMO}, the reliability is guaranteed by making the data transmission rate no less than the amount of data divided by the tolerable delay.
According to this principle, the channel allocation can also be performed as declared in \cite{2019-TCOM-WirelessAccessURLLC}.
An energy efficient optimization of the number of retransmissions, blocklength, and power is proposed in \cite{2018-JSAC-EnergyLatencyTradeoffInURLLC}, where the data transmission reliability is guaranteed in terms of  latency  violation probability.
The resource allocation in \cite{2021-WCL-LatSensitiveUAV} assigns resource blocks and powers for unmanned aerial vehicle assisted networks to balance sum-rate and transmit power while guaranteeing users' transmission reliability in terms of rate outage probability.
In short blocklength regime, the power and blocklength are allocated in \cite{2020-CL-MinmaxDecodingURLLC} to optimize reliability by minimizing the worst-case decoding-error probability of industrial automation wireless links.

Resource allocation with reliability guarantee turns to be quite complicated  when the wireless channel experiences fast fading during a transmission period in mission-critical communications.
In this case, one can utilize the slowly varying large-scale channel fading information to perform resource allocation.
In \cite{2022-TCOM-URLLCedgeNet}, the transmit powers and edge computing decisions are optimized for computation-intensive and time-sensitive services in industrial IoT networks, where short packets are adopted to improve reliability.
To maximize the weighted sum capacity of users subject to energy constraints, the scheme proposed by \cite{2023-TCOM-RAforCellFreeMIMO} jointly optimizes pilot power and payload power for URLLC services in a smart factory.
Uplink and downlink power control method is developed in \cite{2023-TCOM-PAinCellfreeUAV} for mission-critical URLLC in cell-free massive MIMO networks with coexisting ground users and unmanned aerial vehicles.
In the regime of autonomous driving, mode selection, power control, and resource block allocation are jointly optimized in \cite{2020-IoTJ-DRLbasedModeSelForV2X} for the IoV by a deep reinforcement learning based approach, where the reliability of the vehicle-to-vehicle (V2V) links is guaranteed in terms of signal-to-interference-plus-noise ratio (SINR) outage probability.
In \cite{2019-WCL-RAforV2XLargeDeviation}, fixed powers are assigned to the links in the IoV across a time block, where many channel coherence periods, called slots, are contained in a block with a constant large-scale fading.

Although the reliability requirement of the mission-critical links can be guaranteed by the large deviation theory  in \cite{2019-WCL-RAforV2XLargeDeviation}, we can expect more performance gain if power control is conducted in every slot.
To this end, the transceiver may face a dilemma at the initial phase since the future channel is unknown when making instantaneous power allocation decision.
On one hand, the agent can speculate on packet transmission by consuming a small amount of power but with a risk of failing to transmit all packets before deadline.
On the other hand, using a large power may  alleviate burden for subsequent slots but with a cost of power wasting if the channel is bad at the former phase but good at the latter phase.
To address this dilemma, reinforcement learning models are proposed in \cite{2019-TVT-DRLbasedRAforV2V}  and \cite{2019-JSAC-SpectrumSharingMARL} with the users' reliability requirement reflected in the reward design, where  the weighted consumed time is penalized in \cite{2019-TVT-DRLbasedRAforV2V} and  the throughput is awarded in \cite{2019-JSAC-SpectrumSharingMARL}.
It can be observed that such reward design does not exactly match the reliability requirement.
Moreover, it is usually difficult to measure how good the performance can be, especially from an optimization perspective, although they are indeed an effective approach to deal with resource allocation problem in complicated scenarios \cite{2020-Proc-DLwirelessRAwithVNET}.

\subsection{Contributions}

In this paper, we explicitly formulate the sequential power allocation as a  policy optimization problem, aiming to minimize the expected sum power during the transmission period subject to a constraint on the transmission success probability for mission-critical applications.
Due to some good properties held by the policy that maximizes the Lagrangian, we construct a reinforcement learning framework with appropriate reward design such that the goal-directed agent, i.e., the transceiver, obtains the optimal policy that maximizes the Lagrangian by learning to maximize its expected return.
For the model-based case with channel distribution information available at the transceiver, we propose a fast converging algorithm to find the optimal dual solution and its corresponding policy.
For the model-free case, we develop a three-stage procedure, consisting in order of online sampling, offline learning, and online operation, that works efficiently in practical systems.

The main contribution of this work is as follows.
\begin{itemize}
\item We show some good properties of the dual problem of the proposed power allocation policy optimization problem.
\item We design a reinforcement learning framework such that the goal-directed agent exactly maximizes the Lagrangian by maximizing its expected return.
\item We propose a fast converging algorithm to optimize the dual variable for both model-based and model-free cases.
\item We develop a three-stage strategy for practical model-free scenarios, where a backward Q-learning scheme with full exploitation of the sampled channel realizations is proposed to accelerate the learning process.
\end{itemize}

The rest of this paper is organized as follows. Section \ref{Sec:SystemModel} presents the system model and formulates the power allocation as a policy optimization problem.
Afterwards, Section \ref{Section:DualProblem} discusses the dual problem and provides a deep analysis into the structure.
To address the optimization problem, a reinforcement learning framework and the corresponding policy optimization schemes are developed in Section \ref{Section:RLbasedOpt}.
Finally, simulation results are presented in Section \ref{Section:SimuResults} and the conclusion is summarized in Section \ref{Section:Conlusion}.

\section{System Model and Preliminaries}\label{Sec:SystemModel}

In this section, we will introduce the system model considered in this article, formulate the sequential power allocation problem, and present necessary preliminaries of reinforcement learning to be used thereafter.

\subsection{System Model}

Consider a mission-critical network composed of multiple transmitter-receiver pairs or transmission links.
Different links are assumed to occupy mutually orthogonal spectrum to carry  mission-critical messages, which demands performance guarantee on transmission reliability and latency for all links.
This gives rise to an interference-free scenario and we take one link as a representative to formulate a fundamental system without causing performance loss.
Such a single-link model without interference involved moderates the complexity brought about by mutual effect between different links, allows us to extract neat  analytical results,  and serves as a manageable starting point for the problem investigated in this paper.

The channel of the considered link is experiencing the block fading, i.e., the channel power gain remains constant as $h_t$ over the $t$th time slot and is independent and identically distributed (i.i.d.) across different slots.
The time slot here can be regarded as the channel coherence period, which is usually on the order of hundreds of microseconds in a vehicular environment \cite{2019-TWC-RAforVehLowLatHighReliability}.
The system necessitates keeping the transmission outage probability below $\delta$, where the event of transmission outage is defined as that a payload of $N$ data packets are failed to be delivered to the receiver within $T$ slots.
By channel information feedback, the transmitter can be aware of the CSI of the current slot but not the future slots due to the causality.
The main notations used in this article is given in Table \ref{table:Notations}.

\begin{table}
\begin{footnotesize}
\centering
\caption{Table of Notations}\label{table:Notations}
\begin{tabular}   {|m{0.15 \linewidth}|m{0.75\linewidth}|}
\hline
{\textbf{Notation}}        & \textbf{Description} \\\hline
$T$               & Number of slots                                    \\\hline
$N$               & Number of packets                                  \\\hline
$W$               & System bandwidth                                   \\\hline
$Z$               & Number of bits  in each packet                     \\\hline
$\tau$            & Length of each slot                                \\\hline
$\delta$          & Maximum allowed transmission outage probability    \\\hline
$\sigma^2$        & Noise power                                        \\\hline
$h_t$             & Channel power gain in the $t$th slot               \\\hline
$C_t$             & Capacity in the $t$th slot                         \\\hline
$D_t$             & Number of packets transmitted in the $t$th slot    \\\hline
$A_t$             & Transmit power or action in the $t$th slot         \\\hline
$S_t$             & State in the $t$th slot                            \\\hline
$R_t$             & Reward in the $t$th slot                           \\\hline
$U_s$             & Number of slots left before the deadline in state $s$  \\\hline
$V_s$             & Number of packets awaiting transmission in state $s$   \\\hline
$H_s$             & Discretized channel power gain in state $s$            \\\hline
$\mathcal{S}$     & The set of all states                              \\\hline
$\mathcal{S}^{-}$ & The set of all nonterminal states                  \\\hline
$\mathcal{S} \setminus \mathcal{S}^{-}$ & The set of nonterminal states  \\\hline
$\Upsilon({\pi})$ & Transmission success probability of policy $\pi$   \\\hline
$\Psi({\pi})$     & Expected sum power of policy $\pi$                 \\\hline
$\lambda$         & Lagrange multiplier                                \\\hline
$v_{\pi}(s)$      & Value of state $s$ under policy $\pi$                   \\\hline
$q_{\pi}(s,a)$    & Value of state-action pair $(s,a)$ under policy $\pi$   \\\hline
\end{tabular}
\end{footnotesize}
\end{table}

Letting $A_t$ denote the transmit power in the $t$th slot, we have the channel capacity in the $t$th slot given by
\begin{equation}
C_t = W \log_2\left(1 + \frac{h_t A_t}{\sigma^2}\right),
\end{equation}
where $W$ is the frequency bandwidth and $\sigma^2$ is the variance of additive white Gaussion noise (AWGN).
We assume that the transmit power can only take $L$ discrete levels, i.e., selecting from the set $\mathcal{A}=\{a_1, a_2, \cdots, a_L\}$.
The number of packets that can be successfully transmitted to the receiver in the $t$th slot, denoted by $D_t$, has the probability mass function (pmf)
\begin{equation}\label{Eq:channelPacketCapacity}
 {\mathbb{P}}\left\{D_{t}=j\right\}  =   {\mathbb{P}} \left\{ \left\lfloor \frac{ C_{t} \tau }{Z} \right\rfloor = j \right\}, j=0,1,2,\cdots
\end{equation}
where ${\mathbb{P}}\{\cdot\}$ denotes the probability of the event expressed in the braces, $\tau$ is the duration of a slot, $Z$ is the number of bits contained in each data packet, and $\lfloor x \rfloor$ rounds the positive scalar $x$ to the nearest integer towards zero.

Generally, instantaneous channel power gain serves as one of the factors that influence the power selection in the current slot.
However, it takes continuous values and thus necessitates us to distinguish each channel realization from infinite possible cases, which is usually quite complicated.
Fortunately, from \eqref{Eq:channelPacketCapacity}, the channel power gain can be discretized into intervals with break points
\begin{equation}
\hat{h}_{l,j} = \frac{\left(2^{jZ(\tau W)^{-1}} - 1 \right) \sigma^2}{a_l},
\end{equation}
denoting the minimum required channel power gain for carrying $j$ ($j=1,2,\cdots$) packets with transmit power $a_l$ ($l=1,2,\cdots, L$).
This quantization does not prevent us from attaining the optimum performance because the number of carried packets with different transmit powers are identical if the channel power gains fall in the same interval but may vary otherwise.
Moreover, the number of valid intervals will be quite limited since  $\hat{h}_{l,j}$, growing exponentially with $j$, will  become  unreachable rapidly as $j$ increases.
The following example illustrates this quantization method.

\begin{Example}
Consider a system with $\tau=1$ ms, $W=1$ MHz, $Z=8,000$ bits, $\sigma^2=-100$ dBm, $L=3$, $a_1=0$ mW, $a_2=10$ mW, and $a_3=100$ mW.
We have
$\hat{h}_{2,1}=-86.0$ dB, $\hat{h}_{2,2}=-61.8$ dB, $\hat{h}_{2,3}=-37.8$ dB, $\hat{h}_{2,4}=-13.7$ dB, $\hat{h}_{2,5}=10.4$ dB,
$\hat{h}_{3,1}=-96.0$ dB, $\hat{h}_{3,2}=-71.8$ dB, $\hat{h}_{3,3}=-47.8$ dB, $\hat{h}_{3,4}=-23.7$ dB, $\hat{h}_{3,5}=0.41$ dB,
and $\hat{h}_{1,j}=0$ for all $j=1,2,\cdots$.
It can be observed that the channel power gain larger than $\hat{h}_{2,j}$ or $\hat{h}_{3,j}$ for $j>4$ cannot be reached in general in practical situations.
Then, we can discretize the channel power gain into 11 intervals, that are $[0, \hat{h}_{3,1})$, $[\hat{h}_{3,1}, \hat{h}_{2,1})$, $[\hat{h}_{2,1}, \hat{h}_{3,2})$, $[\hat{h}_{3,2}, \hat{h}_{2,2})$, $[\hat{h}_{2,2}, \hat{h}_{3,3})$, $[\hat{h}_{3,3}, \hat{h}_{2,3})$, $[\hat{h}_{2,3}, \hat{h}_{3,4})$, $[\hat{h}_{3,4}, \hat{h}_{2,4})$, $[\hat{h}_{2,4}, \infty)$, without loss of performance.
\end{Example}

\subsection{Problem Formulation}

The transmitter needs to allocate transmit powers for all slots to satisfy the reliability requirement.
While transmission at the highest power level achieves the best possible reliability, it also leads to high energy consumption.
As a result, a judicious power allocation strategy should minimize the power consumption while guaranteeing the data transmission reliability high enough.
This power control problem might be faced by many mission critical tasks.
On one hand, most devices in industrial IoT networks need to reliably transfer control dependent information with strict power constraints \cite{2023-TII-URLLCindIoT}.
On the other hand, energy-limited environment monitoring devices in wireless sensor networks are usually responsible for risk detection and thus necessitates reliable feedback of perception data using power as little as possible \cite{2022-ICC-AnalysisMEC}.
In addition, even in applications with less shortage of energy, such as  safety-critical V2V communications, it is still helpful to investigate the minimum required power for satisfying reliability requirement, which might be further exploited to explore interference management techniques in a spectrum-sharing situation \cite{2022-TCOM-TwoTimescale}.
Therefore, the power allocation strategy we will propose can achieve efficient tradeoff between reliability guarantee and energy consumption for such kinds of mission critical applications.

This gives rise to a functional optimization problem, which aims to find the best mapping from the system states to appropriate power levels.
The state of the system can be fully represented by the combination of three values: the number of slots left before deadline, the number of packets awaiting transmission, and the discretized instantaneous channel condition.
Moreover, a state is defined as a terminal state if there is no slot left in this state and a nonterminal state otherwise.
We use $\mathcal{S}$ to denote the finite set of all states and $\mathcal{S}^{-}$ to denote the finite set of all nonterminal states, allowing us to derive the finite set of terminal states given by  $\mathcal{S} \setminus \mathcal{S}^{-}$.

A power allocation policy in this paper refers to a mapping, denoted by $\pi(s)$, from the set $\mathcal{S}^{-}$ to the set $\mathcal{A}$, i.e., assigning an appropriate power level when the system is in the state $s$.
Under a given power allocation policy $\pi\in {\Pi}$, where $\Pi$ is the set of all possible policies, the expected sum power is
\begin{equation}
\Psi({\pi}) = {\mathbb{E}}_{\pi} \left[ \sum\nolimits_{t=1}^T A_t \right]
\end{equation}
and the transmission success probability is
\begin{equation}
\Upsilon({\pi}) = {\mathbb{P}}_{\pi} \left\{ \sum\nolimits_{t=1}^T D_t \ge N \right\},
\end{equation}
where $\mathbb{E}_{\pi}[\cdot]$ defines the expectation of the random variable in the square brackets  and ${\mathbb{P}}_{\pi}\{\cdot\}$ denotes the probability of the event expressed in the braces, both conditional on $\pi$ being followed.
The power control policy optimization to minimize the expected sum power  while guaranteeing the transmission success probability can be formulated as
\begin{subequations}\label{OptProblem}
\begin{eqnarray}
 \underset{  \pi  }{\min}     \ \ \    & \Psi({\pi})                    \label{OptProblem_obj} \\
{\rm s.t.}                    \ \ \    & \Upsilon({\pi}) \ge 1-\delta.  \label{OptProblem_con}
\end{eqnarray}
\end{subequations}

If there are two different policies, $\pi_1$ and $\pi_2$,  resulting in the same transmission success probability, $\Upsilon(\pi_1)=\Upsilon(\pi_2)$, and the same expected sum power, $\Psi(\pi_1)=\Psi(\pi_2)$, we will not specifically differentiate them for ease of discussion.
Since there are $L$ candidate power levels for each nonterminal state,  the number of possible policies is  $|\Pi|= L^{|\mathcal{S}^{-}|}$, growing exponentially with $T$ and $N$.
Therefore, exhaustive search becomes infeasible in most practical scenarios, which motivates us to design an efficient approach to address the problem \eqref{OptProblem}.

The main difficulty in solving this problem lies in the fact that the future channel fading realizations are unknown by the transmitter.
Furthermore, the transmitter may not even know the distribution of the channel fading in some practical systems.
In this case, the transmitter has almost no knowledge about the best power choice before data transmission, and only after it makes several tries in a state and receives response from the network does it gain some information about consequences of a choice.
A wealth of information about what to do in order to achieve the goal can be derived after a number of trials, which aligns with the framework of reinforcement learning  \cite{2018-Book-ReinforcementLearning}.
However, the constraint in \eqref{OptProblem_con}, which is not typical of a reinforcement learning setting, needs to be properly handled.
To this end, we take a dual domain perspective to the problem \eqref{OptProblem} and derive some nice properties, which shed lights on designing a proper reinforcement learning model.

\subsection{Preliminaries of Reinforcement Learning}

\begin{figure}
\centering
\subfigure[Reinforcement learning framework. \label{fig:RLframework}] {\includegraphics[width=0.46\textwidth]{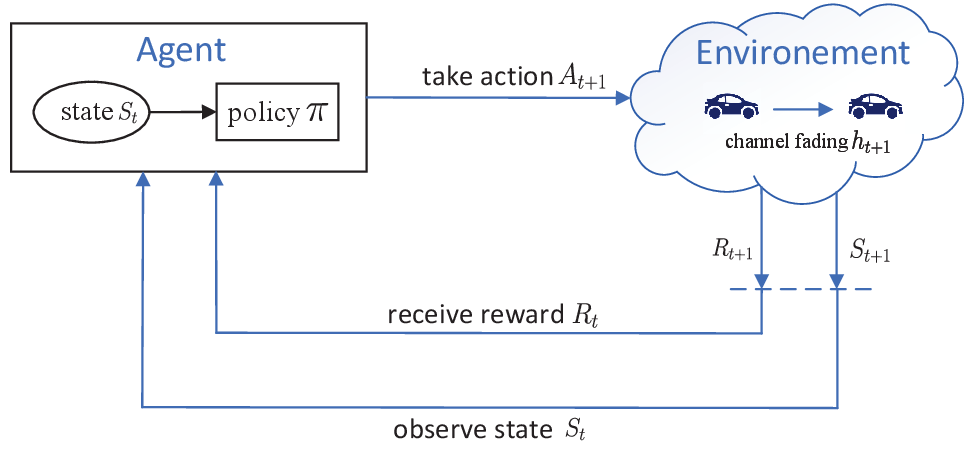}}
\subfigure[Trajectory of agent-environment interaction. \label{fig:RLtrajectory}]{\includegraphics[width=0.46\textwidth]{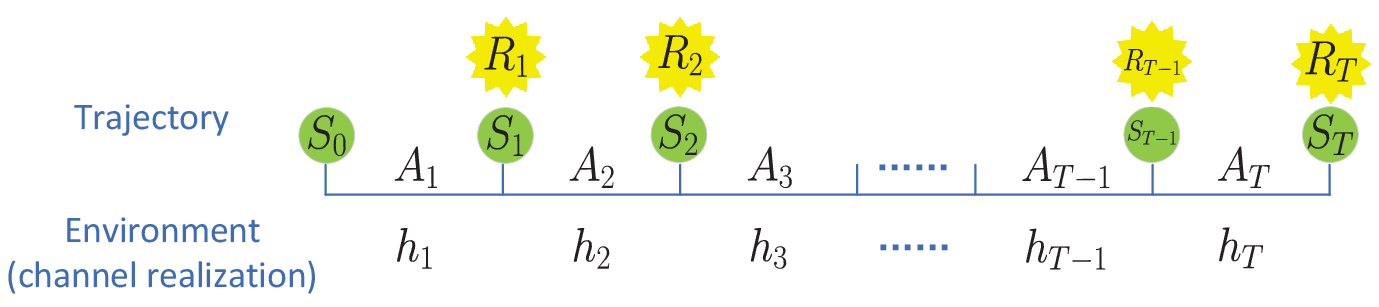}}
\caption{Reinforcement learning framework and its trajectory. \label{fig:RLframework_and_RLtrajectory}}
\end{figure}

In reinforcement learning, the interaction between the agent and environment can be formulated as a  Markov decision process (MDP) as illustrated in Fig. \ref{fig:RLframework}.
In this work, the agent refers to the transceiver and everything beyond the agent is considered as the environment.

In general, during the agent-environment interaction,  the agent observes a state, $S_t$,  at time $t$ from the state space, $\mathcal{S}^{-}$, and on that basis selects an action, $A_{t+1}$, from the action space $\mathcal{A}$, based on a policy, $\pi$.
One time slot later, the environment responds to  action $A_{t+1}$ taken in  state $S_t$ by presenting a new state, $S_{t+1}$, from  state space  $\mathcal{S}$ and giving rise to a reward, $R_{t+1}$, from  reward space, $\mathcal{R}$, to the agent, where the reward is a special scalar value that the agent want to maximize over time \cite{2018-Book-ReinforcementLearning}.
At time $t$, $A_t$ shares the same definition with the one given in Section \ref{Sec:SystemModel}, i.e., representing the power allocation in the $t$th slot,
$R_t$ is the reward received at the end of $t$th slot, and $S_t$ is the state observed at the end of the $t$th slot, for all $t\in \{1,2,\cdots, T\}$.
Because the agent has only $T$ slots to transmit data, the agent-environment interaction leads to a finite trajectory as shown  in Fig. \ref{fig:RLtrajectory}.
It can be observed that the learning task is episodic with each episode starting from  initial state $S_0$ and terminated $T$ slots later.

The MDP is said to be finite if the sets of $\mathcal{S}$, $\mathcal{A}$, and $\mathcal{R}$ have  finite numbers of elements.
Then,  random variables $R_t$ and $S_t$ for $t\in\{1,2,\cdots, T\}$ have discrete probability distributions dependent only on the preceding state and action.
Therefore, we can characterize the dynamics of the MDP by
\begin{equation}
p(s',r|s,a) =  {\mathbb{P}} \{ S_t = s', R_t =r | S_{t-1}=s, A_{t}=a\},
\end{equation}
denoting the probability of state $s'$ and reward $r$ at time $t$ given  preceding state $s$ at time $t-1$ and action $a$ at time $t$, where $s'\in \mathcal{S}$, $r\in \mathcal{R}$, $s\in \mathcal{S}^-$,  and $a\in \mathcal{A}$.

To measure how good a policy is, we define value functions for states and for actions.
In particular, the value function of a state $s$ under a policy $\pi$, denoted by $v_{\pi}(s)$, is the expected return starting from $s$ and following $\pi$ thereafter for all $s\in\mathcal{S}^{-}$, and $V_{\pi}(s)=0$ for all $s\in\mathcal{S}-\mathcal{S}^{-}$.
Similarly, the value of taking action $a$ in state $s$ under policy $\pi$, denoted by $q_{\pi}(s,a)$, as the expected return starting from $s$, taking the action $a$, and thereafter following policy $\pi$, for all $s\in\mathcal{S}^{-}$ and  $a\in\mathcal{A}$,  and $q_{\pi}(s,a)=0$ for all $s\in\mathcal{S}-\mathcal{S}^{-}$ and $a\in\mathcal{A}$.
The functions, $v_{\pi}$ and $q_{\pi}$, are called  state  and action value functions for policy $\pi$, respectively.
Particularly, if a policy $\pi$  makes $v_{\pi}(s)\ge v_{\pi'}(s)$ for all $s\in \mathcal{S}$ for any other policy $\pi'$, $\pi$ is the optimal policy in the reinforcement learning framework.

Dynamic programming leverages value functions to organize and structure the search for the best polices by optimizing the value functions, where the optimal value functions, $v_{*}$, satisfy the Bellman optimality equations
\begin{equation}
\begin{split}
v_{*}(s) & = \underset{a}{\max} \ \mathbb{E}\left[ R_{t+1} + v_{*}(S_{t+1}) | S_t =s, A_{t+1} = a \right] \\
         & = \underset{a}{\max} \ \sum_{s'\in\mathcal{S}} \sum_{r\in\mathcal{R}} p(s',r|s,a) [r + v_{*}(s')]
\end{split}
\end{equation}
for all $s\in\mathcal{S}^{-}$ and $v_{*}(s)=0$ for all $s\in\mathcal{S}-\mathcal{S}^{-}$.
In general, as long as the distribution of the channel fading is known, the agent can figure out  $p(s',r|s,a)$, based on which $\Psi(\pi)$ and $\Upsilon(\pi)$ can be further  derived for any policy $\pi$.
By the value iteration approach  summarized in Algorithm \ref{Algo:PolicyOptGivenLambda} \cite{2018-Book-ReinforcementLearning}, we can efficiently obtain the optimal value function, $v_{*}$, yielding a deterministic action taken in state $s\in\mathcal{S}^{-}$ as
\begin{equation}\label{eq:actionOfPolicy}
\pi(s) = \arg \underset{a}{\max} \ \sum_{s'\in\mathcal{S}} \sum_{r\in\mathcal{R}} p(s',r|s,a) [r + v_{*}(s')].
\end{equation}

\begin{algorithm}[t]
 \caption{Value Iteration for Policy Optimization \cite{2018-Book-ReinforcementLearning}} \label{Algo:PolicyOptGivenLambda}
\begin{algorithmic}[1]
\begin{footnotesize}

\STATE \textbf{Initialization:}

\begin{itemize}
\item  the error tolerance: $\xi$
\item  set $V(s)=0$ for all $s\in\mathcal{S}$
\end{itemize}

\REPEAT

\STATE  $\Delta = 0$

\FOR {$s\in\mathcal{S}^{-}$}

\STATE  $v=V(s)$

\STATE  $V(s) = \max\limits_{a\in\mathcal{A}} \left\{ \sum\limits_{s'\in\mathcal{S}} \sum\limits_{r\in\mathcal{R}} p(s',r|s,a)[r+V(s')] \right\}$

\STATE  $\Delta = \max\{\Delta, |V(s) - v|\}$

\ENDFOR

\UNTIL {$\Delta < \xi$}

\STATE  $v_{*}(s)=V(s)$ for all $s\in\mathcal{S}$

\STATE \textbf{Return:}  policy $\pi$ with $\pi(s)$  given in \eqref{eq:actionOfPolicy}.

\end{footnotesize}
\end{algorithmic}
\end{algorithm}

\section{Dual Problem}\label{Section:DualProblem}

We transform the problem \eqref{OptProblem} to an equivalent form
\begin{subequations}\label{OptProblem2}
\begin{eqnarray}
 \underset{  \pi  }{\max}     \ \ \    & - \ \Psi({\pi}) \\
{\rm s.t.}                    \ \ \    & \Upsilon({\pi}) - (1-\delta) \ge 0,
\end{eqnarray}
\end{subequations}
whose optimal policy is denoted by  $\pi^*$.
By augmenting the objective function with the weighted constraint function, we derive the Lagrangian
\begin{equation}\label{eq:LagranFunction}
 L(\pi, \lambda)  =   - \Psi({\pi}) + \lambda \left[ \Upsilon({\pi})   - (1-\delta)   \right]
\end{equation}
where $\lambda\in[0, +\infty)$ associated with the constraint is called the Lagrange multiplier or the dual variable.
Under any given $\lambda$, we utilize $\pi_{\lambda}$ to denote the power allocation policy maximizing the Lagrangian, i.e.,
\begin{equation}\label{eq:DefPiLambda}
\pi_{\lambda} = \arg \max_{\pi} L(\pi, \lambda) = \arg \max_{\pi} \left\{ - \Psi({\pi})+ \lambda \Upsilon({\pi})\right\},
\end{equation}
implying that $\pi_{\lambda}$ does not depend on $\delta$.
Then, the Lagrange dual function, defined as the maximum value of the Lagrangian over $\pi$, can be expressed as
\begin{equation} \label{eq:dualFunction}
  \ f(\lambda) = \max_{\pi} L(\pi, \lambda), 
\end{equation}
which is always convex since it is the maximum of a family of linear functions of $\lambda$ \cite{2004-Book-ConvexOpt}.
It is intuitive that $\lambda$ plays a role in controlling the tradeoff between power consumption and transmission reliability if the policy, $\pi_{\lambda}$, is implemented.
Generally, a greater $\lambda$ tilts the policy toward transmission reliability.
In particular, only power consumption is taken into account when $\lambda=0$ and only transmission reliability is considered when $\lambda \rightarrow \infty$.

For any given dual variable $\lambda$, the dual function provides an upper bound on the optimal value of the problem \eqref{OptProblem2} \cite{2004-Book-ConvexOpt}.
It is thus straightforward to obtain the best upper bound by minimizing $f(\lambda)$, i.e., addressing the Lagrange dual problem
\begin{subequations}\label{DualProblem}
\begin{eqnarray}
 \underset{  \lambda  }{\min}   \ \ \    &  f(\lambda)= - \Psi({\pi}_{\lambda}) + \lambda \left[ \Upsilon({\pi}_{\lambda})   - (1-\delta)   \right] \\
{\rm s.t.}                      \ \ \    &  \lambda \ge 0,
\end{eqnarray}
\end{subequations}
whose optimal variable and optimal objective value are denoted by $\lambda^*$ and $d^*=f(\lambda^*)$, respectively.
From the duality theory, the weak duality, i.e., $-\Psi(\pi^*) \le d^*$, generally holds even if the original problem is nonconvex \cite{2004-Book-ConvexOpt}.

\subsection{Properties of Policy $\pi_{\lambda}$}
Let us first present the definition of Pareto optimality.
\begin{Definition}[Pareto Optimal Policy]
A policy $\pi$ is said to be a Pareto optimal policy if there is no other policy yielding a higher transmission success probability and at the same time a lower expected sum power.
\end{Definition}

From the definition, Pareto optimality is a situation where no criterion can be better off without making the other criterion worse off.
The following lemma, proved in Appendix \ref{Appendix_Lemma:ParetoOpt}, shows a basic property of $\pi_{\lambda}$.
\begin{Lemma}\label{Lemma:ParetoOpt}
For any given $\lambda\in[0,\infty)$, the  policy $\pi_{\lambda}$ is a Pareto optimal policy.
\end{Lemma}

Generally, if a policy is not Pareto optimal, it is not preferred in practical applications since there is another better policy improving one performance metric without reducing the other.
Based on Lemma \ref{Lemma:ParetoOpt}, we derive the following Lemma, proved in Appendix \ref{Appendix_Lemma:expected sum powertransmission success probabilitymonoto}.
\begin{Lemma}\label{Lemma:expected sum powertransmission success probabilitymonoto}
The transmission success probability, ${\Upsilon}({\pi_{\lambda}})$, and expected sum power, ${\Psi}({\pi_{\lambda}})$, resulting from the policy $\pi_{\lambda}$  are nondecreasing with  $\lambda$.
\end{Lemma}

From Lemmas \ref{Lemma:ParetoOpt} and \ref{Lemma:expected sum powertransmission success probabilitymonoto},  the Pareto optimal policy $\pi_{\lambda}$ produces monotonically increasing transmission success probability and expected sum power as $\lambda$ grows from $0$ to $\infty$.
Nevertheless, not all Pareto optimal policies can be reached by $\pi_{\lambda}$.
To show the reachable and the unreachable Pareto optimal policies, we introduce Lemma \ref{Lemma:ConvexEnvelope1}, proved in Appendix \ref{Appendix_Lemma:ConvexEnvelope1}, and Lemma \ref{Lemma:ConvexEnvelope2}, proved in Appendix \ref{Appendix_Lemma:ConvexEnvelope2}, respectively.

\begin{Lemma}\label{Lemma:ConvexEnvelope1}
Consider two arbitrary Pareto optimal policies, $\pi^b$ and $\pi^c$, satisfying $\Psi({\pi^b}) < \Psi({\pi^c})$ and $\Upsilon({\pi^b}) < \Upsilon({\pi^c})$.
Then, there must exist $\lambda_1$ and $\lambda_2$ ($0\le \lambda_1 < \lambda_2$), such that $\pi_{\lambda_1} = \pi^b$ and $\pi_{\lambda_2} = \pi^c$, if  $\Psi({\pi'})>\theta_{\pi'} \Psi({\pi^b}) + (1-\theta_{\pi'})\Psi({\pi^c})$ holds for any other Pareto optimal policy, $\pi'$, where  real scalar $\theta_{\pi'}$ is the solution to $\Upsilon({\pi'})=\theta_{\pi'} \Upsilon({\pi^b}) + (1-\theta_{\pi'})\Upsilon({\pi^c})$.
\end{Lemma}
\begin{Lemma}\label{Lemma:ConvexEnvelope2}
Consider $0\le\lambda_1<\lambda_2$ such that $\Psi(\pi_{\lambda_1})<\Psi(\pi_{\lambda_2})$ and $\Upsilon(\pi_{\lambda_1})<\Upsilon(\pi_{\lambda_2})$.
Then,  Pareto optimal policy $\pi'$, if it exists, cannot be reached by $\pi_{\lambda}$, if $\Psi(\pi') > \theta_{\pi'} \Psi(\pi_{\lambda_1}) + (1-\theta_{\pi'})\Psi(\pi_{\lambda_2})$, where $\theta_{\pi'}\in (0,1)$ is the solution to $\Upsilon(\pi') = \theta_{\pi'} \Upsilon(\pi_{\lambda_1}) + (1-\theta_{\pi'})\Upsilon(\pi_{\lambda_2})$.
\end{Lemma}

Lemmas \ref{Lemma:ParetoOpt}-\ref{Lemma:ConvexEnvelope2} can be neatly summarized in geometry.
In particular, let us consider a performance figure  with horizontal and vertical axes representing transmission success probability and expected sum power, respectively.
Then, each possible policy can be mapped onto a particular point in the figure.
In this context, by varying $\lambda$ from zero to infinity,  $\pi_{\lambda}$ can  reach all  performance points that are located on the convex envelope of all possible policies, in the order of increasing transmission success probability and expected sum power.
Now, we have a more general conclusion summarized in the following theorem, which can be directly obtained by following Lemmas \ref{Lemma:ParetoOpt}-\ref{Lemma:ConvexEnvelope2}.

\begin{Theorem}\label{Theorem:KeyResult}
Suppose that  $Y$ different Pareto optimal policies, denoted by $\pi^{(1)},\pi^{(2)},\cdots,\pi^{(Y)}$, can be reached by varying $\lambda$ from 0 to $\infty$, where
\begin{equation}
\Psi(\pi^{(1)}) < \Psi(\pi^{(2)}) < \cdots < \Psi(\pi^{(Y-1)})< \Psi(\pi^{(Y)})
\end{equation}
and
\begin{equation}
\Upsilon(\pi^{(1)}) < \Upsilon(\pi^{(2)}) < \cdots < \Upsilon(\pi^{(Y-1)})< \Upsilon(\pi^{(Y)})
\end{equation}
hold.
Then, there must exist
\begin{equation}
0=\lambda^{(0)}<\lambda^{(1)}<\lambda^{(2)}<\cdots<\lambda^{(Y-1)}<\lambda^{(Y)}=\infty,
\end{equation}
such that  $\pi_{\lambda}=\pi^{(j)}$ as long as $\lambda\in[\lambda^{(j-1)}, \lambda^{(j)}]$ for all $j=1,2,\cdots,Y$.
Particularly, the policy $\pi_{\lambda}$ can correspond to either $\pi^{(j)}$ or $\pi^{(j+1)}$ when $\lambda=\lambda^{(j)}$ for $j=1,2,\cdots, Y-1$, where $\lambda^{(j)}$ satisfies
\begin{equation}
\lambda^{(j)}=\frac{\Psi({\pi^{(j+1)}}) - \Psi({\pi^{(j)}})}{\Upsilon({\pi^{(j+1)}}) - \Upsilon({\pi^{(j)}})}.
\end{equation}
\end{Theorem}

From Theorem \ref{Theorem:KeyResult}, as $\lambda$ increases from zero to infinity,  policy $\pi_{\lambda}$ switches when crossing a value in the set $\Lambda=$ $\{\lambda^{(0)}$, $\lambda^{(1)}$, $\lambda^{(2)}$, $\cdots$, $\lambda^{(Y-1)}$, $\lambda^{(Y)}\}$.
However, the fact that both $\pi^{(j)}$ and $\pi^{(j+1)}$ are the optimal policies for $\lambda^{(j)}$ will cause  great inconvenience.
To this end, in the remainder of this paper, the policy $\pi_{\lambda^{(j)}}$ will solely refer to $\pi^{(j+1)}$ that has a higher transmission success probability than $\pi^{(j)}$.

\subsection{Properties of $\pi_{\lambda^*}$}

In the previous subsection, we have shown good properties of $\pi_{\lambda}$, which is independent of $\delta$.
However, the best value of $\lambda$ that minimizes $f(\lambda)$ is dependent on $\delta$.
To derive the optimal $\lambda$ to the dual problem  \eqref{DualProblem}, we first consider the case that the primal power allocation problem is infeasible, i.e., there is no policy leading to the transmission success probability to be no less than $1-\delta$.
In this case, because $f(\lambda)$ is convex and $f'(\lambda)=\Upsilon({\pi}_{\lambda})  - (1-\delta) <0$  for any $\lambda>0$, one can always reduce $f(\lambda)$ by increasing $\lambda$, directly leading to $\lambda^* = \infty$ and $d^*=-\infty$.
This demonstrates that the dual problem is unbounded if the primal problem is infeasible, conforming with the duality theory \cite{2004-Book-ConvexOpt}.
For the feasible case, we have Lemmas \ref{Lemma:OptimalityCase1} and  \ref{Lemma:OptimalityCase2} presented as follows, which are proved in Appendices \ref{Appendix_Lemma:OptimalityCase1} and \ref{Appendix_Lemma:OptimalityCase2}, respectively.

\begin{Lemma}\label{Lemma:OptimalityCase1}
If there exists a $\lambda_0$ such that $\Upsilon(\pi_{\lambda_0})=1-\delta$, then the strong duality holds and the optimal dual variable satisfies  $\Upsilon(\pi_{\lambda^*}) = 1-\delta$.
\end{Lemma}
\begin{Lemma}\label{Lemma:OptimalityCase2}
If the primal problem is feasible and  $\Upsilon(\pi_{\lambda})\ne 1-\delta$ for all $\lambda\in[0,\infty)$, the optimal dual variable satisfies  $\Upsilon(\pi_{\lambda^*}) > 1-\delta$.
Moreover, there is no other $\lambda'$ yeilding $\Upsilon(\pi_{\lambda'}) > 1-\delta$ and $\Psi(\pi_{\lambda'})<\Psi(\pi_{\lambda^*})$.
\end{Lemma}

According to Lemmas \ref{Lemma:OptimalityCase1} and \ref{Lemma:OptimalityCase2}, $\pi_{\lambda^*}$ achieves the best Pareto optimal policy on the convex envelope of all possible policies in the performance figure with horizontal axis being transmission success probability and vertical axis being expected sum power, in the sense that there is no other $\lambda$ giving rise to $\Upsilon(\pi_{\lambda})\ge 1-\delta$ and $\Psi(\pi_{\lambda})<\Psi(\pi_{\lambda^*})$.
Leveraging Lemma \ref{Lemma:expected sum powertransmission success probabilitymonoto} further, we can derive  $\lambda^*$ by addressing an equivalent problem
\begin{subequations}\label{EquiDualProblem}
\begin{eqnarray}
 \underset{  \lambda  }{\min}   \ \ \    &  \lambda \\
{\rm s.t.}                      \ \ \    &  \lambda \ge 0 \\
                                \ \ \    &  \Upsilon(\pi_{\lambda})\ge 1-\delta,
\end{eqnarray}
\end{subequations}
which can be exploited in our algorithm design.

\section{Reinforcement Learning based Optimization}\label{Section:RLbasedOpt}

In this section, we will instantiate the reinforcement learning framework that can be utilized to derive  policy $\pi_{\lambda}$ for an arbitrary given dual variable $\lambda$.
Then, the schemes for deriving the optimal variable $\lambda^*$ are proposed for model-based and model-free cases sequentially.
The main novelty lies in the design of reward to establish the equivalence between maximizing the expected return in reinforcement learning and maximizing Lagrangian in optimization, design a low-complexity iterative algorithm for optimizing the Lagrange multiplier, and propose a fast converging backward Q learning algorithm to address the problem of sparse reward.

\subsection{MDP under Fixed Lagrange Multiplier}\label{subsec:RLframework}
In this subsection, we use  reinforcement learning  to obtain policy $\pi_{\lambda}$ that maximizes the Lagrangian $L(\pi,\lambda)$ over $\pi$ for any given $\lambda$.
As clarified in Section \ref{Sec:SystemModel}, any given state $s\in\mathcal{S}$  should reflect the number of slots left before the deadline in state $s$, $U_{s}$, the number of packets awaiting transmission in state $s$, $V_{s}$, and the discretized channel power gain of the subsequent slot observed in state $s$, $H_{s}$, where $H_{s}$ is discretized from the real continuous channel power gain $h_s$, written as $H_s = \mathfrak{D}(h_{s})$.
Then, the state can be expressed by
\begin{equation}
s=\{U_{s}, V_{s}, H_{s}\},
\end{equation}
In particular, the starting state, $S_0$, always satisfies  $U_{S_0}=T$ and $V_{S_0}=N$.

Towards obtaining policy $\pi_{\lambda}$ that maximizes $L(\pi,\lambda)$ over $\pi$ for a given $\lambda$, we set the reward as
\begin{equation}\label{eq:RewardDesign}
R_{t} = -  A_{t} + \lambda c_{t}
\end{equation}
for $t\in\{1,2,\cdots, T\}$, where  $c_{t}$ is the basic reward for encouraging successful transmission defined as
\begin{equation}
c_{t}=
\begin{cases}
0, & {\rm if} \ \ t < T   \\
\delta - 1, & {\rm if} \ \ t = T \ {\rm and} \ V_{S_t}>0 \\
\delta, & {\rm if} \ \ t = T \ {\rm and} \ V_{S_t}=0.
\end{cases}
\end{equation}
The state value can be figured out as
\begin{equation}
\begin{split}
  & v_{\pi}(s)
=  \ {\mathbb{E}}_{\pi} \left[ \sum_{j=1}^{U_s} R_{T-U_s+j} \middle\vert  S_{T-U_s} =s \right]
\end{split}
\end{equation}
and the state-action value is
\begin{equation}
\begin{split}
  & q_{\pi}(s,a)
=  \ {\mathbb{E}}_{\pi} \left[ \sum_{j=1}^{U_s} R_{T-U_s+j} \middle\vert  S_{T-U_s} =s, A_{T-U_s+1}=a \right].
\end{split}
\end{equation}

A state value table with size $|\mathcal{S}^{-}|\times 1$ and a state-action value table with size $|\mathcal{S}^{-}|\times |\mathcal{A}|$ will be used to organize the policy optimizing process for model-based and model-free cases, respectively.
Since  $|\mathcal{S}^{-}| = M + (T-1)(N+1)M$ and $|\mathcal{A}|=L$, where $M$ is the number of discretized channel power gains, the dimensions of the tables are a polynomial functions of $T, N, L$, and $M$.
The dimensions are acceptable for many practical situations but may face great challenge in the storage of value tables and complexity of training when some of these parameters grow too large.
Although there may exist many optimal policies corresponding to the reinforcement learning task, we use $\pi_{\lambda}^{\dag}$ to denote the specific one maximizing the  transmission success probability, where the subscript $\lambda$ indicates that the optimal policy is influenced by the reward design in  \eqref{eq:RewardDesign}.
The following theorem, proved in Appendix \ref{Appendix_Theorem:RLPerformance}, shows an appealing property of the policy $\pi_{\lambda}^{\dag}$.
\begin{Theorem}\label{Theorem:RLPerformance}
$\pi_{\lambda}^{\dag} = \pi_{\lambda}$ if the dynamics of the MDP are fixed.
\end{Theorem}

Theorem \ref{Theorem:RLPerformance} implies that the optimal policy of the reinforcement learning task maximizes the Lagrangian, $L(\pi, \lambda)$, over $\pi$ for any given $\lambda$, allowing us to solve the primal power allocation problem from the dual perspective by reinforcement learning.
Owing to this, we will use the identical notation, $\pi_{\lambda}$, to represent the optimal policy of the optimization in \eqref{eq:DefPiLambda} and the optimal policy of the reinforcement learning task.
We will first solve the reinforcement learning task under any fixed $\lambda$ and then optimize $\lambda$.

\begin{figure}
\centering
\includegraphics[width=0.40\textwidth]{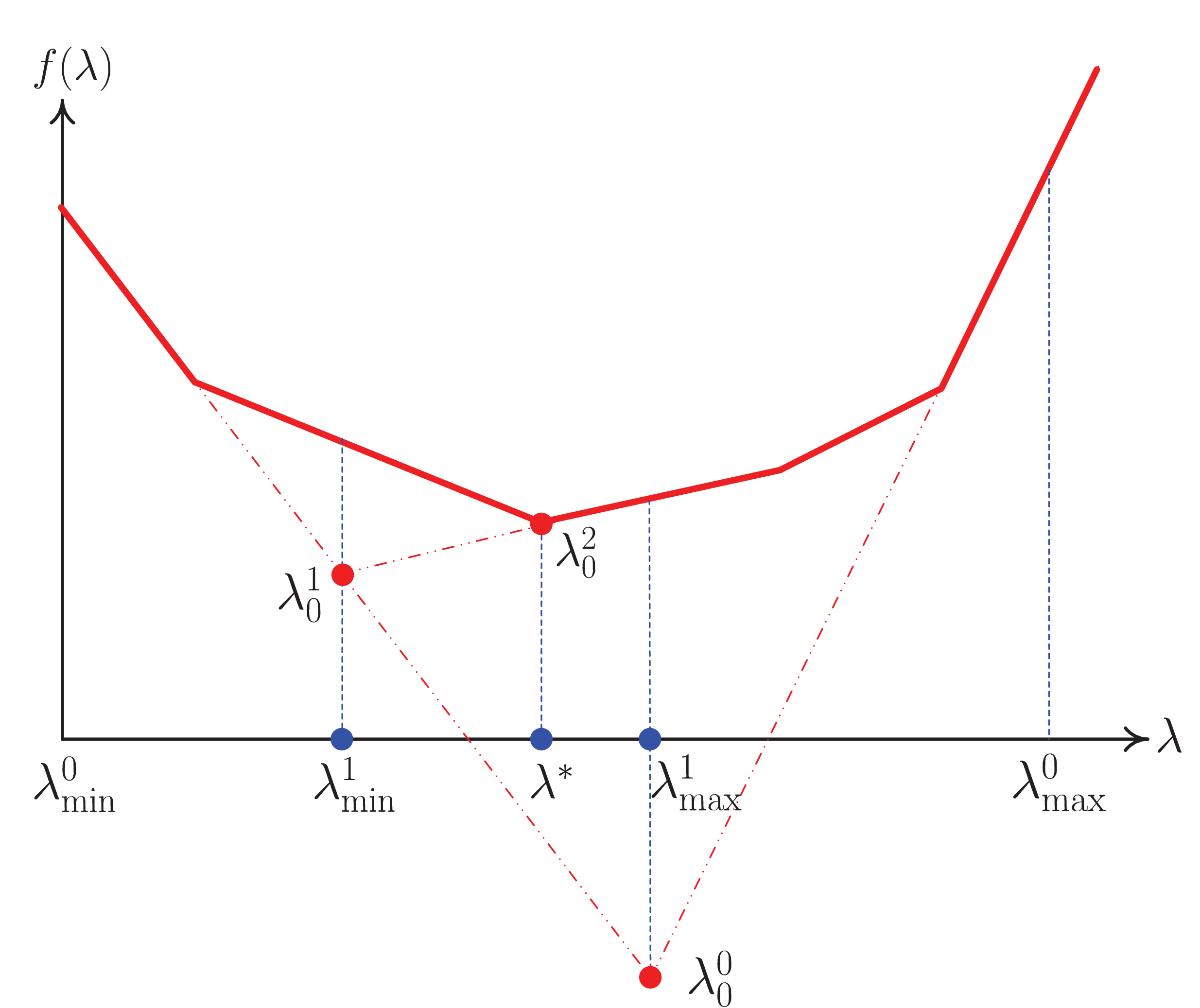}
\caption{Illustration of the trajectory for searching $\lambda^*$, where $[\lambda_{\min}^0, \lambda_{\min}^1]$, $[\lambda_{\max}^0, \lambda_{\max}^1]$, $[\lambda_{0}^0, \lambda_{0}^1, \lambda_{0}^2]$ denote the trajectories of $\lambda_{\min}$, $\lambda_{\max}$, and $\lambda_0$, respectively.
The algorithm starts from $\lambda_{\min}^0$ and $\lambda_{\max}^0$, and then reaches $\lambda_0^0$, $\lambda_{\max}^1=\lambda_0^0$, $\lambda_0^1$, $\lambda_{\min}^1=\lambda_0^1$, $\lambda_0^2$, and $\lambda^*=\lambda_0^2$ in order. \label{Fig_Algo_Explanation}}
\end{figure}

\subsection{Model-Based Approach}\label{Section:ModelBased}

In this subsection, we consider that the agent perfectly knows the environment dynamics characterized by $p(s',r|s,a)$.
The motivation for investigating such a model-based case lies in two aspects.
On one hand, channel statistics might be available in some specific applications, which makes $p(s',r|s,a)$ be known in advance.
On the other hand, the model-based approach provides a high-quality performance baseline to measure strategies in a situation with unknown environment dynamics.

For each $\lambda$, the policy  $\pi_{\lambda}$ can be efficiently derived by Algorithm \ref{Algo:PolicyOptGivenLambda}.
Then, according to Theorem \ref{Theorem:KeyResult}, the convex dual function $f(\lambda)$ is a chain of connected line segments in geometry as shown in Fig. \ref{Fig_Algo_Explanation}, which
instructs us to efficiently search for $\lambda^*$ that minimizes $f(\lambda)$.
Particularly, we set $\lambda_{\min}=0$ and initialize $\lambda_{\max}$ as a sufficient large number.
It is straightforward that $\lambda^*=\lambda_{\min}$ if $\Upsilon(\pi_{\lambda_{\min}})\ge 1-\delta$ and $\lambda^*=\lambda_{\max}$ if $\Upsilon(\pi_{\lambda_{\max}})< 1-\delta$, where the latter corresponds to an infeasible case.
Generally, we have $\Upsilon(\pi_{\lambda_{\min}})< 1-\delta$ and $\Upsilon(\pi_{\lambda_{\max}})> 1-\delta$, indicating $\lambda_{\min}\le\lambda^*\le\lambda_{\max}$ since $\lambda^*$ is the optimal solution the problem \eqref{EquiDualProblem}.
Next, we can look into $\lambda_0$ given by
\begin{equation}\label{eq:lambda0InAlgo}
\lambda_0= \frac{\Psi({\pi_{\lambda_{\max}}}) - \Psi({\pi_{\lambda_{\min}}})}{\Upsilon({\pi_{\lambda_{\max}}}) - \Upsilon({\pi_{\lambda_{\min}}})},
\end{equation}
representing the optimal value of $\lambda$ if there is no $\lambda \in [\lambda_{\min},\lambda_{\max}]$ such that $\pi_{\lambda}\ne \pi_{\lambda_{\min}}$ and $\pi_{\lambda}\ne \pi_{\lambda_{\max}}$. In particular, if
\begin{equation}\label{eq:fdagLambda0}
\begin{split}
 f(\lambda_0) & = - \Psi(\pi_{\lambda_{0}}) + \lambda_0 [\Upsilon(\pi_{\lambda_{0}})-(1-\delta)] \\
              & = - \Psi(\pi_{\lambda_{\max}}) + \lambda_0 [\Upsilon(\pi_{\lambda_{\max}})-(1-\delta)]
\end{split}
\end{equation}
holds, we have $\lambda^* = \lambda_{0}$ and $\pi_{\lambda^*}=\pi_{\lambda_{\max}}$.
Otherwise, there must exist at least one policy $\pi_{\lambda}$ other than ${\pi_{\lambda_{\min}}}$ and ${\pi_{\lambda_{\max}}}$ for $\lambda \in [\lambda_{\min},\lambda_{\max}]$, and we need to further check if there is a better $\lambda$.
To this end, we set $\lambda_{\min}=\lambda_0$ if $\Upsilon(\pi_{\lambda_0})<1-\delta$ and $\lambda_{\max}=\lambda_0$ otherwise, because we need to find the minimum $\lambda$ to satisfy the reliability requirement as elaborated in the problem \eqref{EquiDualProblem}.
Continuing this procedure as illustrated in Algorithm \ref{Algo:ModelBased}, $\lambda^*$ can be found  after fewer than $Y$ iterations, as shown in the exemplary searching trajectory in Fig. \ref{Fig_Algo_Explanation}.

\begin{algorithm}[t]
\caption{Model-Based Algorithm to Solve Problem \eqref{OptProblem2} \label{Algo:ModelBased}}
\begin{algorithmic}[1]
\begin{footnotesize}

\STATE \textbf{Initialization:}

\begin{itemize}
\item  the search range of $\lambda$: $[\lambda_{\min}, \lambda_{\max}]$
\item  the error tolerance: $\theta$
\end{itemize}

\STATE Derive $\pi_{\lambda_{\min}}$, $\Psi(\pi_{\lambda_{\min}})$, and $\Upsilon(\pi_{\lambda_{\min}})$

\STATE Derive $\pi_{\lambda_{\max}}$, $\Psi(\pi_{\lambda_{\max}})$, and $\Upsilon(\pi_{\lambda_{\max}})$

\WHILE  {{\rm True}}

\STATE Set $\lambda_0$ according to \eqref{eq:lambda0InAlgo}

\STATE Derive $\pi_{\lambda_{0}}$, $\Psi(\pi_{\lambda_{0}})$, $\Upsilon(\pi_{\lambda_{0}})$, and $f(\lambda_{0})$

\IF   { $|f(\lambda_0) - \{ - \Psi(\pi_{\lambda_{\max}}) + \lambda_{0} [\Upsilon(\pi_{\lambda_{\max}}) -(1-\delta)] \} | < \theta$ }

\STATE $\lambda^* = \lambda_{0}$ and $\pi_{\lambda^*}=\pi_{\lambda_{\max}}$

\STATE break

\ELSIF {$\Upsilon(\pi_{\lambda_{0}}) < 1-\delta$ }

\STATE  $\lambda_{\min} = \lambda_0$

\ELSE

\STATE  $\lambda_{\max} = \lambda_0$

\ENDIF

\ENDWHILE

\STATE \textbf{Return:} optimal dual variable $\lambda^*$ and optimal policy $\pi_{\lambda^*}$

\end{footnotesize}
\end{algorithmic}
\end{algorithm}

\subsection{Model-Free Approach}

\begin{algorithm}[t]
\caption{Q-Learning Based Policy Optimization} \label{Algo:ModelFree}
\begin{algorithmic}[1]
\begin{footnotesize}

\STATE Initialize the number of sampling episodes: $K$

\vspace{0.5em}

\textbf{$\bullet$ Stage 1: Online Sampling} (done by transmitter unless specified)

\STATE Set $\pi_{\rm sam}$ such that $\pi_{\rm sam}(s)=\max(\mathcal{A})$ for all $s\in\mathcal{S}^{-}$

\STATE $k=0$

\REPEAT

    \STATE  $k = k + 1$

    \STATE  Initialize $S=\{T, N, \mathfrak{D}(h_{1, k}) \}$

    \REPEAT

        \STATE  Take action $A=\pi_{\rm sam}(S)$ and observe $S'$

        \STATE  {$S=S'$}

    \UNTIL {$V_S=0$}

\UNTIL {$k=K$}

\STATE The receiver collects  $h_{t,k}$ for all $t=1,2,\cdots, T$ and $k=1,2,\cdots, K$

\vspace{1em}

\textbf{$\bullet$ Stage 2: Offline Learning} (done by receiver unless specified)

\STATE Initialize $\lambda_{\min}$, $\lambda_{\max}$, and $\theta_f$

\STATE Derive $\pi_{\lambda_{\min}}$, $\Upsilon(\pi_{\lambda_{\min}})$, and $\Psi(\pi_{\lambda_{\min}})$ by Algorithm \ref{Algo:OfflineQlearning}

\STATE Derive $\pi_{\lambda_{\max}}$, $\Upsilon(\pi_{\lambda_{\max}})$, and $\Psi(\pi_{\lambda_{\max}})$ by Algorithm \ref{Algo:OfflineQlearning}

\IF {$\Upsilon(\lambda_{\max}) < 1-\delta$}

    \STATE $\lambda^* = \lambda_{\max}$, $\pi_{\lambda^*}=\pi_{\lambda_{\max}}$

\ELSE

\WHILE  {{\rm True}}

\STATE Set $\lambda_0$ according to \eqref{eq:lambda0InAlgo}

\STATE Get $\pi_{\lambda_{0}}$, $\Upsilon(\pi_{\lambda_{0}})$, $\Psi(\pi_{\lambda_{0}})$, and $f({\lambda_{0}})$ by  Algorithm \ref{Algo:OfflineQlearning}

\IF  { $\lambda_0 \ge \lambda_{\max} $  or $\lambda_0 \le \lambda_{\min}$}

    \STATE $\lambda^* = \lambda_{\max}$, $\pi_{\lambda^*}=\pi_{\lambda_{\max}}$

    \STATE break

\ELSIF  { $ \{ f(\lambda_0) - \{ - \Psi(\pi_{\lambda_{\max}}) + \lambda_{0} [\Upsilon(\pi_{\lambda_{\max}}) -(1-\delta)] \}  < \theta_f \}$ }

    \STATE $\lambda^* = \lambda_{0}$, $\pi_{\lambda^*}=\pi_{\lambda_{\max}}$

    \STATE break

\ELSIF {$\Upsilon(\pi_{\lambda_{0}})   < 1-\delta$ }

    \STATE  $\lambda_{\min} = \lambda_0$, $\pi_{\lambda_{\min}}=\pi_{\lambda_{0}}$

    \STATE $\Psi(\pi_{\lambda_{\min}}) = \Psi(\pi_{\lambda_{0}})$, $\Upsilon(\pi_{\lambda_{\min}}) = \Upsilon(\pi_{\lambda_{0}})$

\ELSIF {$\Upsilon(\pi_{\lambda_{0}}) \ge 1-\delta $ }

    \STATE  $\lambda_{\max} = \lambda_0$, $\pi_{\lambda_{\max}}=\pi_{\lambda_{0}}$

    \STATE  $\Psi(\pi_{\lambda_{\max}}) = \Psi(\pi_{\lambda_{0}})$, $\Upsilon(\pi_{\lambda_{\max}}) = \Upsilon(\pi_{\lambda_{0}})$

\ENDIF

\ENDWHILE

\ENDIF

\STATE the receiver feed $\pi_{\lambda^*}$ back to the transmitter and this stage ends

\STATE the transmitter conducts power allocation using policy $\pi_{\rm sam}$ in this stage

\vspace{1em}

\textbf{$\bullet$ Stage 3: Online Operation} (done by transmitter)

\STATE The transmitter conducts power allocation using policy $\pi_{\lambda^*}$

\end{footnotesize}
\end{algorithmic}
\end{algorithm}

If the agent is not aware of the environment model, i.e., the distribution of channel fading is unknown by the agent, the only way to optimize the power allocation policy is to learn from experience.
An exemplary strategy is the off-policy Q-learning, which approximates the best action value function by continuously updating its estimate, $q(s,a)$, in exploiting the experience.
There is a fundamental dilemma in designing an online algorithm to derive the optimal power allocation policy in practical applications.
On one hand, a large number of sample trajectories of channel realizations are required to learn the action value and test the corresponding policy's performance for a given $\lambda$, not to mention the optimization of $\lambda$.
On the other hand, the practical system may not tolerate a lower transmission success probability even during the learning period, making the choice of $\lambda$ difficult during the learning period.
To address this dilemma, we propose  a three-stage algorithm with coordination between the transmitter and the receiver as summarized in Algorithm \ref{Algo:ModelFree}, composed in sequential of the online sampling, offline learning,  and online operation.

\vspace{0.5em}
\noindent $\bullet$ \textbf{Stage 1: Online Sampling}
\vspace{0.5em}

In this initial stage, composed of the first $K$ episodes, the transmitter always utilizes the maximum transmit power in all states, i.e., adopting policy $\pi_{\rm sam}$ such that $\pi_{\rm sam}(s)=\max\{\mathcal{A}\}$ for all $s\in\mathcal{S}^{-}$.
The receiver records the channel power gain in the $t$th slot during the $k$th episode as $h_{t, k}$ in this stage.
It is worth noting that this stage embraces the highest achievable transmission success probability, which overly guarantees the transmission reliability in general.
Rather than performing policy optimization, the most important thing  in this stage is to collect sufficient information about channel dynamics at the receiver, which can be exploited to learn the best power allocation policy in the next stage.
Although highest power consumption is caused in every slot during the first $K$ episodes, the proposed backward Q-learning algorithm can usually converge fast, e.g., after thousands of episodes, as clarified in the simulation results.
This implies that the online sampling stage only takes a small fraction of time compared with the network operation time and thus the additional power consumption in the online sampling stage is acceptable.

\vspace{0.5em}
\noindent $\bullet$ \textbf{Stage 2: Offline Learning}
\vspace{0.5em}

\begin{algorithm}[t]
\caption{Backward Q-Learning and Policy Testing} \label{Algo:OfflineQlearning}
\begin{algorithmic}[1]
\begin{footnotesize}

\STATE \textbf{Initialization:}

\begin{itemize}
\item  dual variable  $\lambda$ and minimum learning rate $\alpha_{\min}$
\item  $q(s,a)=0$ for all $s\in\mathcal{S}$ and $a\in\mathcal{A}$
\item  $k=0$
\end{itemize}

\REPEAT

    \STATE  $k = k + 1$

    \STATE  $\alpha = \max\{{1}/{k}, \alpha_{\min}\}$

    \STATE  $u = 0$

    \REPEAT

        \STATE $u = u+1$

        \FOR {$s\in\{s|U_s=u, H_s = \mathfrak{D}(h_{T-u+1, k}) \}$}

            \FOR {$a \in \mathcal{A}$}

                \STATE  Experience channel fading $h_{T-u+1,k}$, observe next state $s'$,  and receive reward $r$

                \begin{small}
                \STATE  $q(s,a)=q(s,a)+\alpha[r + \max\limits_{a'\in\mathcal{A}} q(s',a') - q(s,a)]$
                \end{small}

            \ENDFOR

         \ENDFOR

     \UNTIL{$u=T$}

\UNTIL {$k=K$}

\STATE $q_{*}(s,a)=q(s,a)$ for all $s\in\mathcal{S}$ and $a\in\mathcal{A}$

\STATE Get policy $\pi_{\lambda}$ with $\pi_{\lambda}(s)= \arg\max\limits_{a\in\mathcal{A}}q_{*}(s,a)$ for $s\in\mathcal{S}^{-}$

\STATE Take policy $\pi_{\lambda}$ for the $K$ episodes, derive the expected sum power $\Psi(\pi_{\lambda})$ estimated by average sum power of the $K$ episodes, and obtain the transmission success probability $\Upsilon(\pi_{\lambda})$ estimated by the ratio of the number of episodes with successful data transmission to $K$

\STATE Set $f(\lambda)=-\Psi(\pi_{\lambda}) + \lambda[\Upsilon(\pi_{\lambda})-(1-\delta)]$

\STATE \textbf{Return:} $\pi_{\lambda}$, $\Psi(\pi_{\lambda})$, $\Upsilon(\pi_{\lambda})$, and $f(\lambda)$

\end{footnotesize}
\end{algorithmic}
\end{algorithm}

In this stage, the transmitter keeps implementing the same strategy with the sampling stage until the end of this stage.
On the receiver side, the receiver can approximate the optimal action value, $q_{*}{(s,a)}$, for any given $\lambda$ by Q-learning with full exploration of channel realizations $h_{t, k}$ for all $t=1,2,\cdots, T$ and $k=1,2,\cdots, K$, leading to the policy $\pi_{\lambda}$ with
\begin{equation}\label{eq:actionOfPolicy_QL}
\pi_{\lambda}(s) = \arg \underset{a}{\max} \  q_{*}{(s,a)}.
\end{equation}
The full exploration means that we can update $q(s,a)$ for all related state-action pairs  once in every episode.
However, the reward accounting for successful transmission is sparse, making the learning still converge slowly.
To this end, we propose a backward Q-learning scheme, where  $q(s,a)$ is updated in an increasing order of $U_s$, i.e., the number of slots left before the deadline in state $s$.
In particular, in the $k$th episode, we first consider a state $s$ with $U_s=1$ and $H_s=\mathfrak{D}(h_{T, k})$, meaning that we start from the beginning of the last slot.
By looking into an action $a$, experiencing the channel fading $h_{T, k}$, we can figure out the next state $s'$ and  the reward $r$, allowing us to update $q(s,a)$ according to
\begin{equation}
q(s,a)=q(s,a)+\alpha[r + \max\limits_{a'\in\mathcal{A}} q(s',a') - q(s,a)],
\end{equation}
where $\alpha=\max\{1/k, \alpha_{\min}\}$ makes $q(s,a)$ approach sample-average return of the state-action pair $(s,a)$ but with a learning rate being at least $\alpha_{\min}$.
It can be observed that the reward information about successful transmission has been broadcast to all states in $\{s|U_s=1, H_s=\mathfrak{D}(h_{T, k})\}$ if we have updated  $q(s,a)$ once for all $s$ in this set and all $a\in\mathcal{A}$.
By continuing this updating for states in $\{s|U_s=2, H_s=\mathfrak{D}(h_{T-1, k})\}$, $\{s|U_s=3, H_s=\mathfrak{D}(h_{T-2, k})\}$, $\cdots$, $\{s|U_s=T, H_s=\mathfrak{D}(h_{1, k})\}$ in order, we can make the reward of successful transmission prorogate to all related state-action values backward in each episode.
After updating the state-action values by leveraging the sampled $K$ episodes of channel realization, we can expect that good approximation of optimal state-action values can be achieved as long as $K$ is not too small.
Then, we test the derived policy over the $K$ episodes and obtain its Monte-Carlo based transmission success probability and expected sum power.
The detailed algorithm of offline Q-learning for deriving $\pi_{\lambda}$ and policy testing is summarized in Algorithm \ref{Algo:OfflineQlearning}.

\begin{Remark}
The Q-learning based method in Algorithm 4 holds significant advantage in obtaining $\pi_{\lambda}$: 1) Low complexity is required since only each value in the Q table needs to be updated once in every episode; 2) Fast learning can be expected using the proposed  backward Q-learning mechanism; 3) Convergence of the training process is guaranteed in general \cite{2001-ConvergenceQlearning}.
To address the dimensionality issues faced by Q-learning in training $\pi_{\lambda}$, one can also leverage deep Q network (DQN) based reinforcement learning methods, where state-action values  are expressed by deep neural networks (DNNs) rather than a table.
However, the DQN based approach may have slower learning speed, greater training complexity, and less guarantee of convergence than the proposed backward Q learning method as shown in simulation results, implying that they are difficult to serve as basis for optimizing the Lagrange multiplier $\lambda$. Therefore, we mainly focus on Q learning in this work.
\end{Remark}

We can search for the optimal dual variable $\lambda$ and optimal policy $\pi_{\lambda}$ after a finite number of iterations by a similar procedure as in the model-based algorithm.
However, the proposed backward Q-learning with full exploitation of channel realizations may still result in suboptimal policies caused by insufficient channel exploitations.
Moreover, inaccurate transmission success probability and expected sum power may be derived in policy testing over the sampled channel realizations, which may cause unintended updating of $\lambda$.
This issue can be easily addressed by modifying the procedure for searching $\lambda$.
In particular, the algorithm immediately returns a policy satisfying reliability requirement when the stopping criterion is met or the search trajectory goes beyond expectation, as shown in Lines 22-27 in Algorithm \ref{Algo:ModelFree}.
In fact, this modification may lead to inaccurate solutions of $\lambda^{*}$, but the derived policy $\pi_{\lambda^{*}}$ is quite efficient.
In other words, the ultimate goal of the algorithm is to find $\pi_{\lambda^{*}}$ rather than $\lambda^{*}$.
Finally, once the optimal policy $\pi_{\lambda^*}$ is derived, it is fed back to the transmitter and the offline learning stage ends.
It is worth noting that the receiver collects the channel realization directly in the first stage and then learns to obtain the optimal policy, $\pi_{\lambda^*}$, by itself in the second stage, which needs no signaling exchange between the transmitter and the receiver. The only singling overhead is the feedback of the optimal policy from the receiver to the transmitter after the second stage.
Because a policy maps from states to actions, the optimal policy, expressed by $|\mathcal{S}^{-}|$ values,  needs to be fed back from the receiver to the transmitter only once, which causes low signaling overhead in general.

\vspace{0.5em}
\noindent $\bullet$ \textbf{Stage 3: Online Operation}
\vspace{0.5em}

In this final stage, the transmitter always takes action using the policy $\pi_{\lambda^*}$, which should achieves the best Pareto optimal solution on the convex envelope of all possible policies in the performance figure with the horizontal and vertical axes being transmission success probability and expected sum power, respectively.

\section{Simulation Results}\label{Section:SimuResults}

\begin{table}
\begin{footnotesize}
\centering
\caption{Simulation Parameters}\label{table:simulaton_para}
\begin{tabular}   {|m{0.55 \linewidth}|m{0.35\linewidth}|}
\hline
{\textbf{Parameter}}       & \textbf{Value} \\\hline
Number of slots $T$        & 6$\sim$40            \\\hline
Number of packets $N$      & 10$\sim$50           \\\hline
Noise power $\sigma^2$     & $-80$ dBm     \\\hline
Action space $\mathcal{A}$ & $\{0, 10, 100\}$ mW \\\hline
Large-scale fading         & $10^{-7}$      \\\hline
Slot length $\tau$         & $1$ ms       \\\hline
Bandwidth $W$              & $1$ MHz \\ \hline
Number of bits  in each packet $Z$     & $2,000$  \\\hline
Number of sampling episodes $K$        & $10^5$ \\\hline
Minimum learning rate                                 & $0.001$     \\\hline
Error tolerance $\xi$, $\theta$, and $\theta_f$       & $10^{-8}$       \\\hline
Initialization of $[\lambda_{\min}, \lambda_{\max}]$  & $[0, 10^{5}]$   \\\hline
Small-scale fading                               & Rayleigh fading \\\hline
\end{tabular}
\end{footnotesize}
\end{table}

In this section, we provide numerical results to validate the analysis and the proposed power allocation strategy.
In particular, we model the channel power gain as the product of large-scale fading and small-scale fading, where the former keeps constant during the payload delivery period and the latter follows exponential distribution with unit mean in every slot.
The setup of fixed large-scale fading is reasonable because it usually does not change too much for hundreds of milliseconds even under high-mobility vehicular environment, while the latency of safety-related payload transmission can be required to be less than tens of milliseconds \cite{2019-TVT-MyTVT2019}.
All simulation parameters are set by default to the values in Table \ref{table:simulaton_para}, whereas each figure may have particular settings taking precedence wherever applicable.

\begin{figure}
\centering
\includegraphics[width=0.48\textwidth]{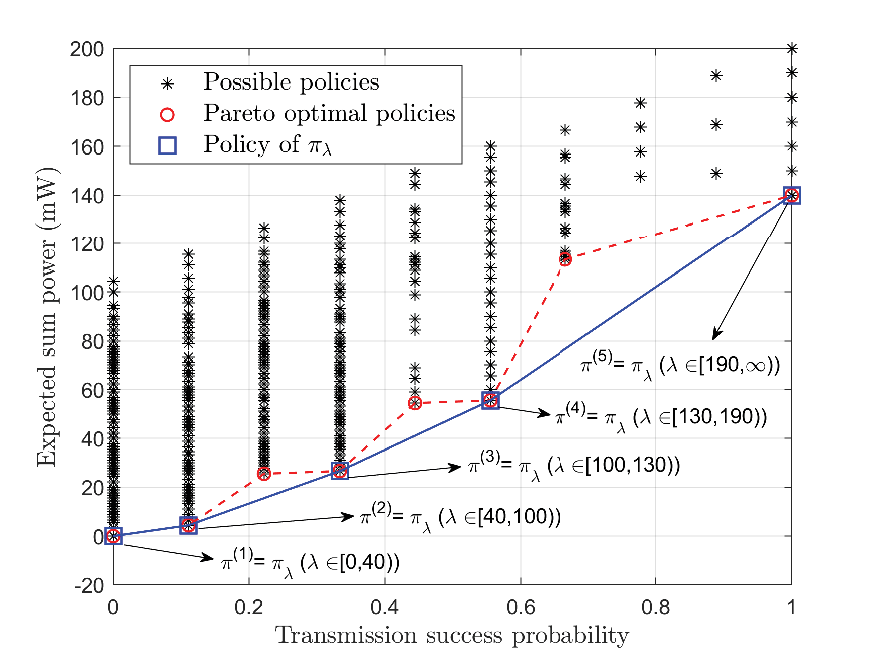}
\caption{The transmission success probability and expected sum power of all possible policies, Pareto optimal policies, and policies $\pi_{\lambda}$ for all $\lambda\in [0,\infty)$. \label{fig:ParetoEnveIllus}}
\end{figure}

Consider a wireless transmission link requiring that $N=4$ packets should be delivered within $T=2$ slots, where  $Z=7,600$ bits, $\sigma^2=10^{-10}$ dBm, and the large-scale fading is $10^{-6}$.
To make the policy search tractable, the small-scale fading can only take values of 0.2 and 8, with probability $2/3$ and $1/3$, respectively.
In this scenario, the number of nonterminal states is  $|\mathcal{S}^{-}|=12$ and the number of possible policies is $|\Pi|= L^{|\mathcal{S}^{-}|}=3^{12}=531,441$.
Leveraging the distribution of channel fading, we can figure out the state transition probability and state distribution probability for each possible policy, based on which the expected sum power and transmission success probability of each policy can be derived.
Then, each policy can be mapped onto a point, marked by an asterisk, in Fig. \ref{fig:ParetoEnveIllus}.
Among all policies, the Pareto optimal ones are marked by red circles and connected by red dashed lines.
Finally, all points derived by $\pi_{\lambda}$, of the total number of 5, are marked by blue squares and are represented  in the order from left to right by $\pi^{(1)}$, $\pi^{(2)}$, $\cdots$, $\pi^{(5)}$, which are connected in order by blue solid lines.
Here, for any $\lambda$, $\pi_{\lambda}$ can be exhaustively searched across all possible policies according to \eqref{eq:DefPiLambda}.
By figuring out $\pi_{\lambda}$ for all $\lambda$, we derive the value set of $\lambda$ yielding $\pi_{\lambda}=\pi^{(i)}$ for $i=1,2,\cdots,5$ as marked in Fig. \ref{fig:ParetoEnveIllus}.
As shown in Fig. \ref{fig:ParetoEnveIllus}, by varying $\lambda$ from zero to infinity, policy $\pi_{\lambda}$ is always Pareto optimal, which verifies Lemma \ref{Lemma:ParetoOpt}.
Moreover, the fact that an increasing $\lambda$ leads to nondecreasing expected sum power and transmission success probability is in line with Lemma \ref{Lemma:expected sum powertransmission success probabilitymonoto}.
Finally, $\pi_{\lambda}$ can reach all the Pareto optimal points on the convex envelope of all possible policies and cannot touch any  point above the convex envelope, which confirms Lemma \ref{Lemma:ConvexEnvelope1} and Lemma \ref{Lemma:ConvexEnvelope2}.

\begin{figure}
\centering
\includegraphics[width=0.48\textwidth]{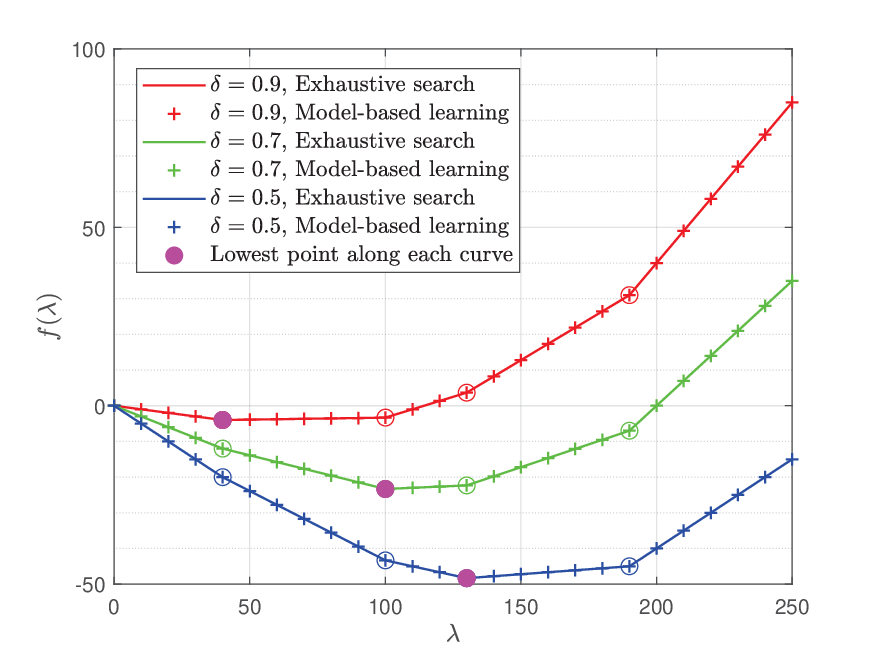}
\caption{Curve of the Lagrange dual function $f(\lambda)$ with respect to $\lambda$, where each line segment of the curves represents a Pareto optimal policy, each hollow or solid circle denotes a policy changing point  along the curves, and each solid magenta circle denotes the lowest point of the curves.  \label{fig:f_lambda}}
\end{figure}

In the same network setting with Fig. \ref{fig:ParetoEnveIllus}, we also show the curves of the dual function, $f(\lambda)$, with respect to $\lambda$ for different values of $\delta$ in Fig. \ref{fig:f_lambda}.
We can utilize the proposed model-based value iteration to figure out the optimal policy that solves the proposed reinforcement learning task, leading to the value of $f(\lambda)$.
Moreover, the exhaustive search can be applied to find policy $\pi_{\lambda}$ that maximizes the Lagrangian as long as $\lambda$ and $\delta$ are given, which provides the true best value of $f(\lambda)$.
It can be observed that the model-based reinforcement learning reaches exactly the optimal policy, $\pi_{\lambda}$, for all $\lambda$, which confirms  Theorem \ref{Theorem:RLPerformance}.
We notice that each line segment between adjacent circles represents a reachable Pareto optimal policy, which always maximizes the Lagrangian along the line segment.
In addition, comparing Fig. \ref{fig:ParetoEnveIllus} with Fig. \ref{fig:f_lambda},  it can be observed that the optimal dual variable $\lambda^*$, i.e., the minimizer of $f(\lambda)$, indeed offers the best policy, $\pi_{\lambda^*}$, on the convex envelope of all policies.

\begin{figure}
\centering
\subfigure[Experienced return for each learning episode. \label{fig:RealReturnVSepisode}]{\includegraphics[width=0.48\textwidth]{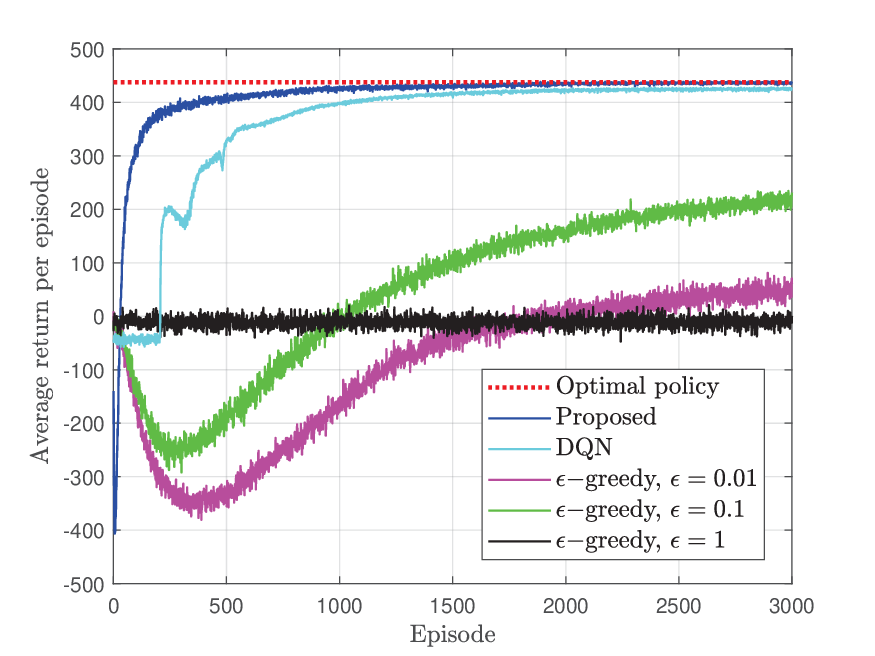}}
\subfigure[Expected return of learned policy after each episode. \label{fig:ExpectedReturnVSepisode}] {\includegraphics[width=0.48\textwidth]{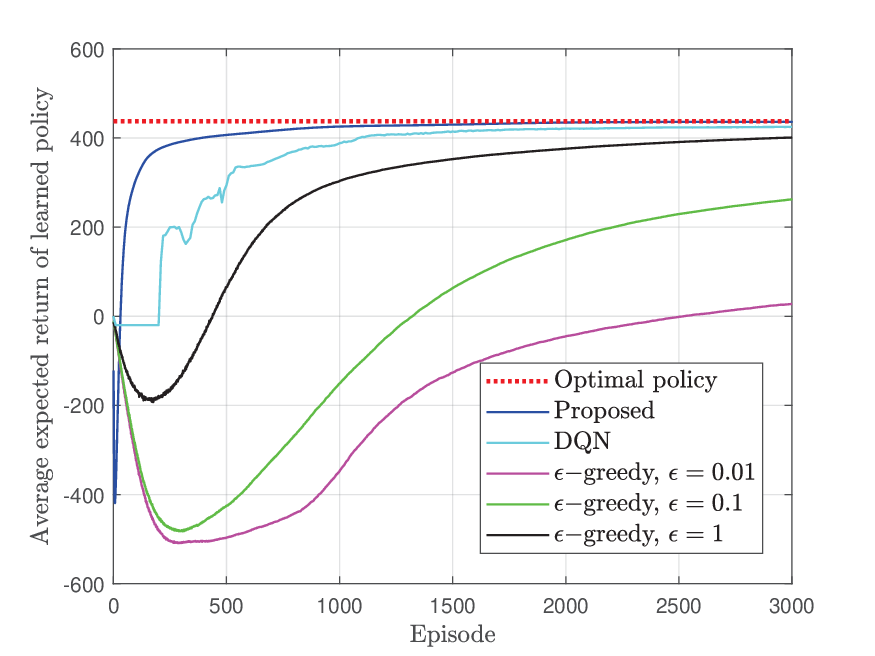}}
\caption{Experienced return and expected return of the learned policy for each episode, where $\lambda=1,000$ and $N=16$ packets are required to be transmitted within $T=10$ slots. \label{fig:ExpectedReturn_and_RealReturn}}
\end{figure}

By conducting $1,000$ independent experiments, Fig. \ref{fig:RealReturnVSepisode} and Fig. \ref{fig:ExpectedReturnVSepisode} show the average actual return of the adopted policy in each episode and the expected return of the learned policy after each episode, respectively.
Since our proposed backward Q-learning scheme is implemented offline,  we consider  $K$ episodes of channel realizations have been sampled in each experiment.
Then, the return of the optimal policy and the proposed method in Fig. \ref{fig:RealReturnVSepisode} are derived by testing the optimal policy and our learned policy  on the same channel realizations of the $\epsilon-$greedy Q-learning, where the optimal policy can be obtained by model-based learning.
Compared with the actual return during the learning process, Fig. \ref{fig:ExpectedReturnVSepisode} shows the average expected return of the policy that greedily chooses the action possessing the highest state-action value in every station according to the learned Q table in each episode.
In particular, we also consider DQN based approach, where the policy training does not begin until 2048 four-tuples of experience, i.e., $(S_t, A_t, R_t, S_{t+1})$,  has been sampled.
Both Fig. \ref{fig:RealReturnVSepisode} and Fig. \ref{fig:ExpectedReturnVSepisode} demonstrate that the proposed backward Q-learning scheme converges fast to the optimal policy within about 1,500 episodes, whereas the traditional $\epsilon-$greedy Q-learning strategy requires more than 3,000 episodes to converge.
Although DQN based method converges faster than $\epsilon-$greedy Q-learning, the convergence to the optimal policy is not guaranteed, implying that the derived policy may deviate from $\pi_{\lambda}$  that we wonder.
We will hence not consider DQN based approach in the optimization of Lagrange multiplier $\lambda$ thereafter.
In addition, $\epsilon-$greedy Q-learning faces the dilemma of trading off exploration and exploitation. On one hand, the random policy, i.e., $\epsilon=1$, converges fastest to the optimal state-action value, but the actual performance during the learning process is quite bad.
On the other hand, $\epsilon=0.1$ or $\epsilon=0.01$ has better return during the learning procedure, but the state-action values need much more episodes to approach the optimum.
The fast convergence of the proposed model-free learning profits from two facts.
First, the backward learning makes the reward information about successful transmission propagate to every related state-action pair in each episode.
Second, the Q value updating for all state-action pairs in each episode fully exploits the sampled channel realizations.

\begin{figure}
\centering
\includegraphics[width=0.48\textwidth]{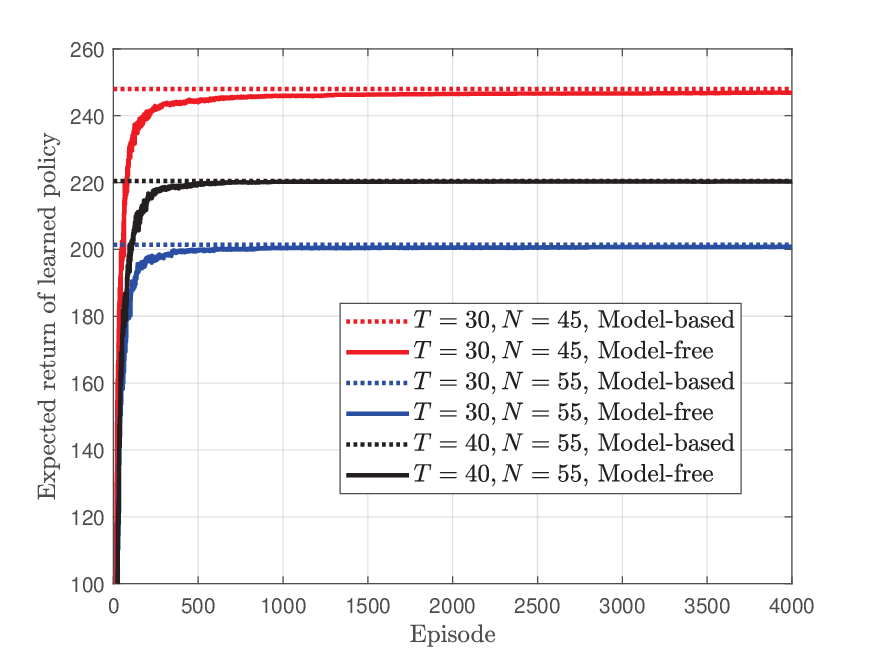}
\caption{Expected return of the learned policy in each episode, where $\lambda=800$.  \label{fig:Fig_LearningPerformance}}
\end{figure}

Fig. \ref{fig:Fig_LearningPerformance} shows the expected return of the learned policy versus episodes with different payloads and time constraints in a randomly selected experiment.
The model-based method provides the theoretical optimal policy.
For the model-free case, we update the policy after each episode  and figure out its expected return according to model-based analysis.
We observe that the expected return of the derived policy in the model-free cases approaches the optimum only by thousands of episodes for different $T$ and $N$, implying that the learning performance is not very sensitive
to the values of $T$ and $N$.
Moreover, when $N$ is reduced, the power consumption is allowed to decrease while guaranteeing the same transmission success probability, which makes the expected return improve.
Instead, the expected return increases with $T$ since less power can be spent to speculate on channel realizations to obtain the same transmission reliability.

\begin{figure}
\centering
\subfigure[Transmission outage probability. \label{fig:Fig_topIteration}]{\includegraphics[width=0.48\textwidth]{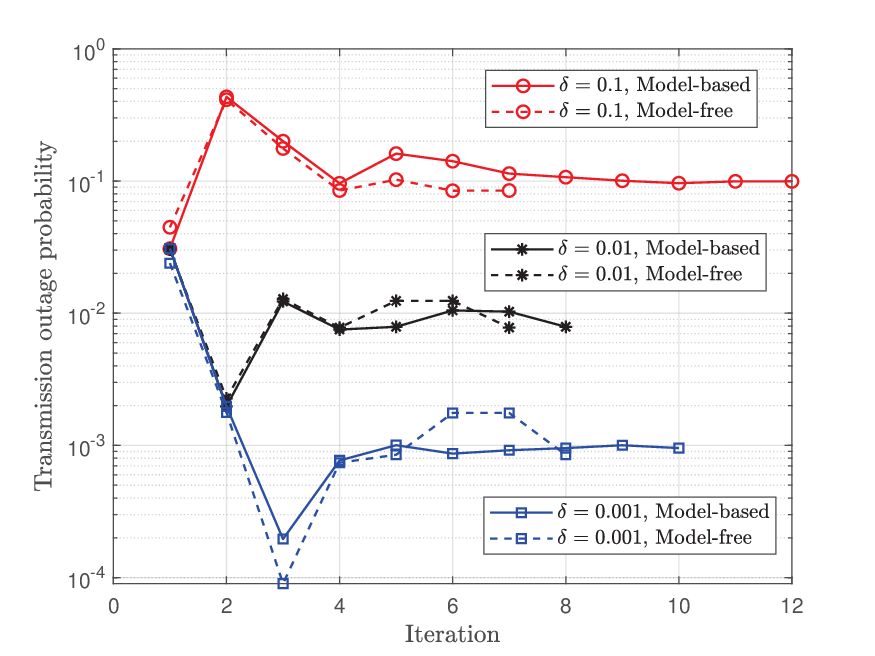}}
\subfigure[Expected sum power. \label{fig:Fig_espIteration}] {\includegraphics[width=0.48\textwidth]{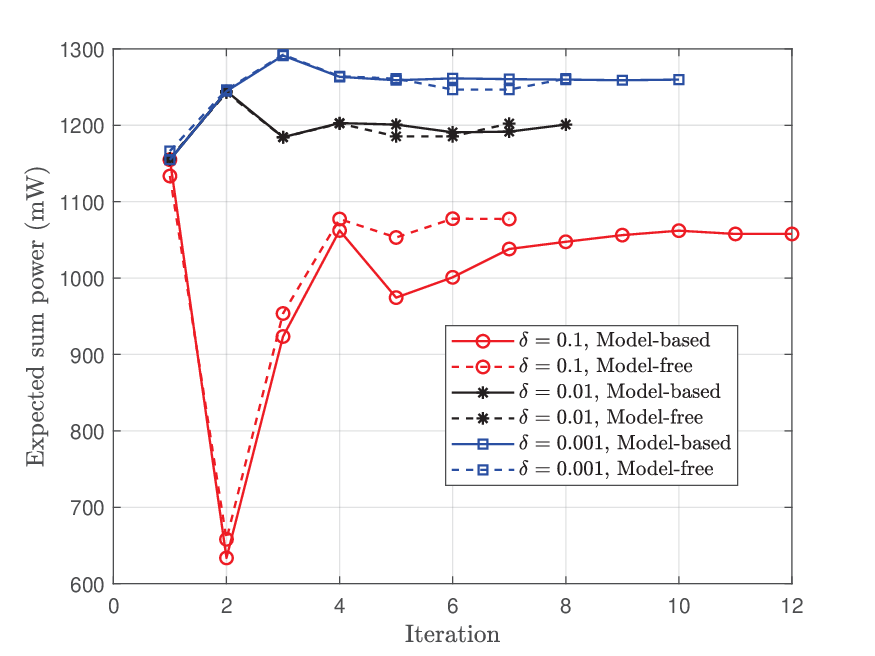}}
\caption{Transmission outage probability and expected sum power versus iteration of $\lambda$, where $N=45$ packets need to be transmitted within $T=30$ slots and $Z=3,500$ bits are contained in each packet. \label{fig:top_esp_iteration}}
\end{figure}

By performing a randomly chosen experiment, the transmission outage probability and expected sum power versus the iteration of $\lambda$ for different transmission outage probability thresholds are shown in Fig. \ref{fig:top_esp_iteration}.
According this figure, it takes only serval iterations to search for the optimal dual variable $\lambda$ for both the model-based and model-free cases.
The final derived policy results in a transmission outage probability close to $\delta$ but satisfying the transmission success probability constraint for different choices of $\delta$, which confirms the effectiveness of the obtained policy.
Moreover, the model-based and model-free cases may differ in their trajectories of the transmission outage probability and expected sum power during the process of searching $\lambda$, but they will finally converge to the policies with quite close performance while satisfying the reliability constraint.
Comparing Fig. \ref{fig:Fig_topIteration} and Fig. \ref{fig:Fig_espIteration}, we find that there is a tradeoff between the transmission outage probability and expected sum power during the iteration of $\lambda$, i.e., improving one of them usually decreases the other.
On the whole, the updating of $\lambda$ tries to make a good balance between transmission outage probability and expected sum power while satisfying the reliability requirement.

\begin{figure}
\centering
\includegraphics[width=0.48\textwidth]{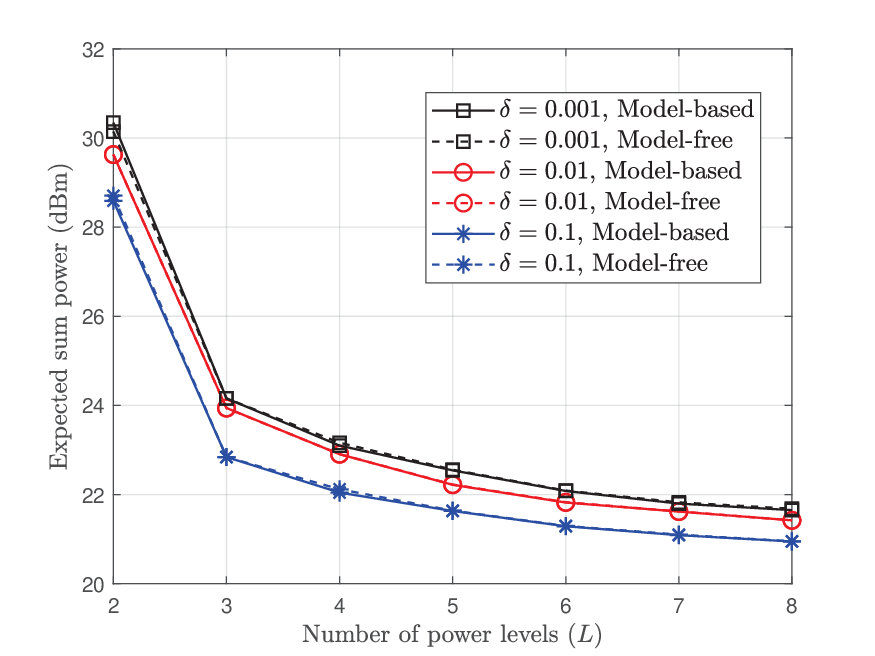}
\caption{Expected sum power with varying number of  power levels with  $T=10$ slots and $N=26$ packets. \label{fig:Fig5_ESPvsPowerLevels}}
\end{figure}

Fig. \ref{fig:Fig5_ESPvsPowerLevels} shows the average expected sum power with different numbers of power levels, $L$, by conducting multiple independent experiments.
The power levels are equally spaced between $10$ dBm and $30$ dBm according to the number of power levels, e.g., the power space is $\mathcal{A}=\{10$ dBm, $20$ dBm,  $30$ dBm$\}$ when $L=3$.
The expected sum power of the model-based case can be figured out according to our analysis, whereas the expected sum power of the model-free case is derived by testing the final learned policy over the sampled channel realizations.
It can be observed that the proposed strategy in the model-free case has quite close expected sum power performance to the optimal model-based case, demonstrating the effectiveness of our proposal in practical scenarios.
Because increasing $L$ provides more  power choices in all states, allowing the transceiver to control power allocation more delicately with smaller granularity,
the expected sum power decreases with $L$.
However, marginal gain is expected as $L$ grows large, indicating that discrete power allocation  with a large $L$ may get close to the performance of continuous power allocation in practical systems.
Moreover, since the reliability constraint becomes more stringent  when we decreases the maximum allowed transmission outage probability $\delta$, more powers are needed to satisfy the reliability requirement and thus the expected sum power increases.

\begin{figure}
\centering
\includegraphics[width=0.48\textwidth]{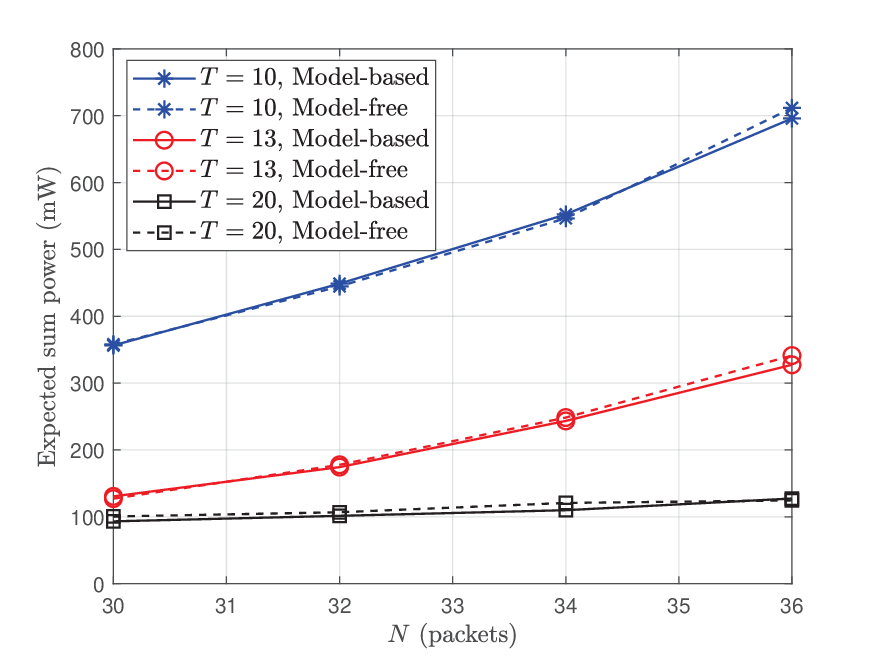}
\caption{Expected sum power under varying $N$, where the maximum allowable transmission outage probability is $\delta=0.1$. \label{fig:Fig4_ESPvsNandT}}
\end{figure}

The average expected sum power with different numbers of packets, $N$, is shown in Fig. \ref{fig:Fig4_ESPvsNandT}, where  the average is made across $100$ independent experiments.
From the figure, the performance of the proposed model-free case matches the model-based case, implying that the proposed algorithm can be applied in practical scenarios.
Moreover, as the payload $N$ increases, the transmitter needs more power the maintain the required reliability, giving rise to the increase of expected sum power.
On the other hand, the transmission task becomes less urgent as $T$ grows, allowing the transmitter to spend less power speculating on channel realizations and thus making the expected sum power decrease.

\section{Conclusion} \label{Section:Conlusion}

Considering a given amount of packets are required to be transmitted within a certain time constraint, this paper investigated power allocation policy optimization aiming to minimize the expected sum power subject to the transmission success probability constraint for mission-critical applications.
We developed a reinforcement learning framework, where the agent can learn to achieve the best policy that maximizes the Lagrangian, which is always Pareto optimal.
A fast converging algorithm is designed to optimize the dual variable for both the model-based and model-free cases.
We also proposed a three-stage procedure for practical networks, consisting of online sampling, offline learning, and online operation.
Based on our simulation results, the proposed reinforcement learning based algorithm in the model-based case perfectly finds the optimal policy of the dual solution and the  model-free method achieves close performance to the model-based method.
It is worth noting that the considered problem would become quite complicated when multiple links share the same spectrum due to the coupling effect of powers taken by different links.
To this end, our future work will take into consideration of interference in a multi-link network to minimize the system power consumption while providing satisfactory payload delivery reliability for all connections, which might be solved by multi-agent reinforcement learning.

\appendices

\section{Proof of Lemma \ref{Lemma:ParetoOpt}}\label{Appendix_Lemma:ParetoOpt}
Consider a particular $\lambda$ and its corresponding optimal policy $\pi_{\lambda}$, leading to the expected sum power and transmission success probability denoted by $\Psi({\pi_{\lambda}})$ and $\Upsilon({\pi_{\lambda}})$, respectively.
We will prove the conclusion by contradiction.
Let us suppose there is another policy, $\pi'$, such that $\Psi({\pi'}) < \Psi({\pi_{\lambda}}) $ and $\Upsilon({\pi'}) \ge \Psi({\pi_{\lambda}})$ hold, or $\Psi({\pi'}) \le \Psi({\pi_{\lambda}})$ and $\Upsilon({\pi'}) > \Psi({\pi_{\lambda}})$ hold.
From the definition of Lagrangian in \eqref{eq:LagranFunction}, we have
\begin{equation}
\begin{split}
    L(\pi', \lambda) & = - \Psi({\pi'}) + \lambda \left[ \Upsilon({\pi'})   - (1-\delta)   \right] \\
                     & > - \Psi({\pi_{\lambda}}) + \lambda \left[ \Upsilon({\pi_{\lambda}})   - (1-\delta)   \right] \\
                     & = \ L(\pi_{\lambda}, \lambda),
\end{split}
\end{equation}
which indicates that the policy $\pi_{\lambda}$ is not the maximizer of $L(\pi, \lambda)$ over $\pi$.
This conflicts with the definition of $\pi_{\lambda}$ and thus the proof is complete.

\section{Proof of Lemma \ref{Lemma:expected sum powertransmission success probabilitymonoto}}\label{Appendix_Lemma:expected sum powertransmission success probabilitymonoto}
Considering  $0\le \lambda_1 < \lambda_2 <+\infty$, we need to show ${\Upsilon}({\pi_{\lambda_1}})\le {\Upsilon}({\pi_{\lambda_2}})$ and ${\Psi}({\pi_{\lambda_1}})\le {\Psi}({\pi_{\lambda_2}})$.
In the following, we will prove this Lemma by contradiction.

Suppose ${\Upsilon}({\pi_{\lambda_1}}) > {\Upsilon}({\pi_{\lambda_2}})$ holds.
According to Lemma \ref{Lemma:ParetoOpt}, we have ${\Psi}({\pi_{\lambda_1}}) > {\Psi}({\pi_{\lambda_2}})$; otherwise the policy $\pi_{\lambda_2}$ is dominated by $\pi_{\lambda_1}$ and is not Pareto optimal.
Noticing the fact that $\pi_{\lambda_1}$ is the maximizer of $L(\pi, \lambda_1)$ over $\pi$, we have $L(\pi_{\lambda_2}, \lambda_1) \le L(\pi_{\lambda_1}, \lambda_1)$, i.e.,
\begin{equation}\label{eq:lambdaPolicyChange1}
\begin{split}
    & \ - {\Psi}({\pi_{\lambda_2}}) + \lambda_1 \left[ {\Upsilon}({\pi_{\lambda_2}}) -(1-\delta)\right] \\
\le & \ - {\Psi}({\pi_{\lambda_1}}) + \lambda_1 \left[ {\Upsilon}({\pi_{\lambda_1}})-(1-\delta)\right].
\end{split}
\end{equation}
Similarly, we derive $L(\pi_{\lambda_1}, \lambda_2) \le L(\pi_{\lambda_2}, \lambda_2)$, i.e.,
\begin{equation}\label{eq:lambdaPolicyChange2}
\begin{split}
    & \ - {\Psi}({\pi_{\lambda_1}}) + \lambda_2 \left[ {\Upsilon}({\pi_{\lambda_1}}) -(1-\delta)\right] \\
\le & \ - {\Psi}({\pi_{\lambda_2}}) + \lambda_2 \left[ {\Upsilon}({\pi_{\lambda_2}})-(1-\delta)\right].
\end{split}
\end{equation}
Jointly processing \eqref{eq:lambdaPolicyChange1} and \eqref{eq:lambdaPolicyChange2}, we have
\begin{equation}
\begin{split}
    & \ \lambda_1 \left[{\Upsilon}({\pi_{\lambda_2}}) - {\Upsilon}({\pi_{\lambda_1}})\right]  \le  {\Psi}({\pi_{\lambda_2}}) - {\Psi}({\pi_{\lambda_1}}) \\
\le & \ \lambda_2  \left[{\Upsilon}({\pi_{\lambda_2}}) - {\Upsilon}({\pi_{\lambda_1}})\right],
\end{split}
\end{equation}
which gives rise to $\lambda_1 \ge \lambda_2$.
This is a contradiction and the proof can be finished.
\setcounter{equation}{41}
\begin{table*}
\begin{equation}\label{eq:ColumnSpan}
\begin{split}
  & \ \max\{ L(\pi_{\lambda_1}, \lambda),  L(\pi_{\lambda_2}, \lambda) \} - L(\pi', \lambda) \\
= & \ \max\left\{ - \Psi(\pi_{\lambda_1}) + \lambda \left[\Upsilon(\pi_{\lambda_1}) - (1-\delta) \right], - \Psi(\pi_{\lambda_2}) + \lambda \left[\Upsilon(\pi_{\lambda_2}) - (1-\delta) \right] \right\} -  [ - \Psi({\pi'}) + \lambda(\Upsilon({\pi'})-(1-\delta))] \\
= & \  \max\left\{ -[\Psi({\pi_{\lambda_1}}) - \Psi({\pi'})] + \lambda[\Upsilon({\pi_{\lambda_1}}) - \Upsilon({\pi'})] ,
                   -[\Psi({\pi_{\lambda_2}}) - \Psi({\pi'})] + \lambda[\Upsilon({\pi_{\lambda_2}}) - \Upsilon({\pi'})]  \right\}\\
> & \  \max\{ -[\Psi({\pi_{\lambda_1}}) - \theta_{\pi'} \Psi(\pi_{\lambda_1}) - (1-\theta_{\pi'})\Psi(\pi_{\lambda_2})]
                                  + \lambda[\Upsilon({\pi_{\lambda_1}}) - \theta_{\pi'} \Upsilon(\pi_{\lambda_1}) - (1-\theta_{\pi'})\Upsilon(\pi_{\lambda_2})], \\
  & \ \ \ \ \ \ \  -[\Psi({\pi_{\lambda_2}}) - \theta_{\pi'} \Psi(\pi_{\lambda_1}) - (1-\theta_{\pi'})\Psi(\pi_{\lambda_2})]
                                  + \lambda[\Upsilon({\pi_{\lambda_2}}) - \theta_{\pi'} \Upsilon(\pi_{\lambda_1}) - (1-\theta_{\pi'})\Upsilon(\pi_{\lambda_2})]  \}\\
= & \ \max\{(1-\theta_{\pi'})[ -(\Psi(\pi_{\lambda_1}) - \Psi(\pi_{\lambda_2})) + \lambda(\Upsilon(\pi_{\lambda_1}) - \Upsilon(\pi_{\lambda_2})) ],
                \theta_{\pi'}[ -(\Psi(\pi_{\lambda_2}) - \Psi(\pi_{\lambda_1})) + \lambda(\Upsilon(\pi_{\lambda_2}) - \Upsilon(\pi_{\lambda_1})) ] \}  \\
> & \ 0
\end{split}
\end{equation}
\center{\rule{18cm}{0.75pt}}
\end{table*}

\setcounter{equation}{32}
\section{Proof of Lemma \ref{Lemma:ConvexEnvelope1}}\label{Appendix_Lemma:ConvexEnvelope1}
We choose the value of $\lambda_0$ as the solution to $L(\pi^b, \lambda) = L(\pi^c, \lambda)$ with respect to $\lambda$, i.e.,
\begin{equation}
- \Psi({\pi^b}) + \lambda[\Upsilon({\pi^b}) - (1-\delta)] = - \Psi({\pi^c}) + \lambda[\Upsilon({\pi^c}) - (1-\delta)],
\end{equation}
which yields
\begin{equation}
\lambda_0=\frac{\Psi({\pi^c}) - \Psi({\pi^b})}{\Upsilon({\pi^c}) - \Upsilon({\pi^b})}.
\end{equation}
Leveraging the fact of $\Psi({\pi'})>\theta_{\pi'} \Psi({\pi^b}) + (1-\theta_{\pi'})\Psi({\pi^c})$ and $\Upsilon({\pi'})=\theta_{\pi'} \Upsilon({\pi^b}) + (1-\theta_{\pi'})\Upsilon({\pi^c})$, we can derive
\begin{equation}\label{eq:Lag1and2forLambda0}
\begin{split}
  & \ L(\pi^b, \lambda_0) - L(\pi', \lambda_0) = L(\pi^c, \lambda_0) - L(\pi', \lambda_0) \\
= & \ \{ - \Psi({\pi^c}) + \lambda_0 \left[ \Upsilon({\pi^c}) - (1-\delta) \right] \} \\
  & \ \ \ \ \ \ \ \ \ \ \ \ \ \ \ \ \ \ \ \ \ \ \ \ \  - \{ - \Psi({\pi'}) + \lambda_0[\Upsilon({\pi'})-(1-\delta)] \} \\
= & \  -[\Psi({\pi^c}) - \Psi({\pi'})] + \lambda_0[\Upsilon({\pi^c}) - \Upsilon({\pi'})] \\
= & \ Y_{\pi'}  -[\Psi({\pi^c}) - \theta_{\pi'} \Psi({\pi^b}) - (1-\theta_{\pi'})\Psi({\pi^c})] \\
  & \ \ \ \ \ \ \ \ \ \ \ \ \ \ \ \ \ \   + \lambda_0[\Upsilon({\pi^c}) - \theta_{\pi'} \Upsilon({\pi^b}) - (1-\theta_{\pi'})\Upsilon({\pi^c})] \\
= & \ Y_{\pi'}  + \theta_{\pi'}\left[ -(\Psi({\pi^c}) - \Psi({\pi^b})) + \lambda_0 (\Upsilon({\pi^c}) - \Upsilon({\pi^b}))\right] \\
= & \ Y_{\pi'},
\end{split}
\end{equation}
where $Y_{\pi'}$ is given by
\begin{equation}
Y_{\pi'} = \Psi({\pi'}) - [\theta_{\pi'} \Psi({\pi^b}) + (1-\theta_{\pi'})\Psi({\pi^c})] > 0,
\end{equation}
for any Pareto optimal policy $\pi'$  other than $\pi^b$ and $\pi^c$.
This implies that $\pi_{\lambda_0}$ must be the policy $\pi^b$ or $\pi^c$ since both $\pi^b$ and $\pi^c$ maximize the Lagrangian $L(\pi, \lambda_0)$.
Then, there must exist a positive $\Delta{\lambda}_2$ such that
\begin{equation}
\theta_{\pi'}\left[ -(\Psi({\pi^c}) - \Psi({\pi^b})) + (\lambda_0+\Delta{\lambda}_2) (\Upsilon({\pi^c}) - \Upsilon({\pi^b}))\right] > -Y_{\pi'}
\end{equation}
holds for any Pareto optimal policy $\pi'$  other than $\pi^b$ and $\pi^c$ since
\begin{equation}
\theta_{\pi'}\left[ -(\Psi({\pi^c}) - \Psi({\pi^b})) + \lambda_0 (\Upsilon({\pi^c}) - \Upsilon({\pi^b}))\right] > 0
\end{equation}
holds for any Pareto optimal policy $\pi'$  other than $\pi^b$ and $\pi^c$.
Following \eqref{eq:Lag1and2forLambda0}, we have
\begin{equation} \label{eq:pi2betterOthers}
L(\pi^c, \lambda_0 + \Delta{\lambda}_2) > L(\pi', \lambda_0 + \Delta{\lambda}_2)
\end{equation}
for any Pareto optimal policy $\pi'$  other than $\pi^b$ and $\pi^c$.
On the other hand,  it is not difficult to obtain
\begin{equation}\label{eq:pi2betterpi1}
\begin{split}
  & \ L(\pi^c, \lambda_0 + \Delta{\lambda}_2) >  L(\pi^b, \lambda_0 + \Delta{\lambda}_2).  \\
\end{split}
\end{equation}
Jointly considering \eqref{eq:pi2betterOthers} and \eqref{eq:pi2betterpi1} , we can conclude there exists $\lambda_2 = \lambda_0+\Delta{\lambda}_2$ such that $\pi_{\lambda_2}=\pi^c$.
Similarly, it can be shown that there must exist $\lambda_1 = \lambda_0-\Delta{\lambda}_1$ such that $\pi_{\lambda_1}=\pi^b$, where $\Delta{\lambda}_1$ is a positive scalar.

\section{Proof of Lemma \ref{Lemma:ConvexEnvelope2}}\label{Appendix_Lemma:ConvexEnvelope2}
For any $\lambda>0$, we have
\begin{equation}
\max\{ L(\pi_{\lambda_1}, \lambda),  L(\pi_{\lambda_2}, \lambda) \} - L(\pi', \lambda) > 0,
\end{equation}
where the detailed deduction can be found in \eqref{eq:ColumnSpan} on the top of this page.
This tells us that the policy $\pi'$ is dominated by either $\pi_{\lambda_1}$ or $\pi_{\lambda_2}$ from the perspective of maximizing the Lagrangian $L(\pi, \lambda)$ over $\pi$ for any $\lambda$, implying that there is no $\lambda$ leading to $\pi_{\lambda}=\pi'$.

\section{Proof of Lemma \ref{Lemma:OptimalityCase1}}\label{Appendix_Lemma:OptimalityCase1}
Without loss of generality, let us  consider that $\lambda_0 \in [\lambda^{(j)}, \lambda^{(j+1)})$, where $j\in \{0,1,\cdots, Y-1\}$.
On one hand,   all $\lambda\in[\lambda^{(j)},\lambda^{(j+1)})$ result in the same value of function
\setcounter{equation}{42}
\begin{equation}
\begin{split}
  & f(\lambda) = - \Psi({\pi}_{\lambda}) + \lambda \left[ \Upsilon({\pi}_{\lambda})  - (1-\delta)   \right] \\
= & - \Psi({\pi}^{(j+1)}) + \lambda \left[ \Upsilon({\pi}^{(j+1)})  - (1-\delta)   \right]\\
= & \Psi({\pi}^{(j+1)}).
\end{split}
\end{equation}
On the other hand,  for any $\lambda' \notin [\lambda^{(j)},\lambda^{(j+1)})$, we have
\begin{equation}
\begin{split}
  &  f(\lambda') = - \Psi({\pi}_{\lambda'}) + \lambda' \left[ \Upsilon({\pi}_{\lambda'})  - (1-\delta)   \right] \\
> & - \Psi({\pi}_{\lambda}) + \lambda' \left[ \Upsilon({\pi}_{\lambda})  - (1-\delta)   \right]  \\
= & - \Psi({\pi}_{\lambda}) + \lambda \left[ \Upsilon({\pi}_{\lambda})  - (1-\delta)   \right]
= f(\lambda),
\end{split}
\end{equation}
where $\lambda\in[\lambda^{(j)},\lambda^{(j+1)})$.
Therefore, the optimal dual solution $\lambda^*$ makes $\Upsilon(\pi_{\lambda^*}) = 1-\delta$ hold, where $\lambda$ can be any value in $[\lambda^{(j)},\lambda^{(j+1)})$.
Moreover, since $\pi_{\lambda^*}$ is a feasible policy, we derive the strong duality, i.e., $-\Psi(\pi^*) = d^*$, resulting from
\begin{equation}
\begin{split}
-\Psi({\pi}^*) & \le d^*  = f(\lambda^*)= - \Psi({\pi}_{\lambda^*}) + \lambda^* \left[ \Upsilon({\pi}_{\lambda^*})  - (1-\delta)   \right] \\
               &  = - \Psi({\pi}_{\lambda^*}) \le - \Psi({\pi}^*).
\end{split}
\end{equation}

\setcounter{equation}{46}
\begin{table*}
\begin{equation}\label{eq:vPiS0eqLPiLambda_deduction}
\begin{split}
  & v_{\pi}(s_0)  = {\mathbb{E}}_{\pi} \left[ \sum_{t=1}^{T} R_{t}(\pi) \right]
=  \lim_{K\rightarrow \infty} \frac{1}{K} \sum_{k=1}^{K} \sum_{t=1}^T R_{t,k}(\pi)
=  \lim_{K\rightarrow \infty}\frac{1}{K} \sum_{k=1}^{K} \sum_{t=1}^T  \left(-  A_{t,k}(\pi) + \lambda c_{t,k}(\pi) \right)\\
= & \lim_{K\rightarrow \infty}\left[ \frac{1}{K} \sum_{k=1}^{K}  \sum_{t=1}^T -  A_{t,k}(\pi) \right] + \lambda \cdot \lim_{K\rightarrow \infty}\left[  \frac{1}{K} \sum_{k=1}^{K}  \sum_{t=1}^T  c_{t,k}(\pi) \right] \\
= & \lim_{K\rightarrow \infty}\left[ \frac{1}{K} \sum_{k=1}^{K}  \sum_{t=1}^T -  A_{t,k}(\pi) \right] + \lambda \cdot \lim_{K\rightarrow \infty}\left[  \frac{1}{K} \sum_{k=1}^{K}  \left[ \delta I_k(\pi)  + (\delta -1)  (1-I_k(\pi)) \right] \right] \\
= & \lim_{K\rightarrow \infty}\left[ \frac{1}{K} \sum_{k=1}^{K}  \sum_{t=1}^T -  A_{t,k}(\pi) \right] + \lambda \cdot \lim_{K\rightarrow \infty}\left[   \delta \cdot \frac{\sum_{k=1}^{K} I_k(\pi)}{K}   + (\delta-1) \cdot \left(1- \frac{\sum_{k=1}^{K} I_k(\pi)}{K} \right) \right]\\
= & -{\Psi}(\pi) + \lambda \left[\delta \Upsilon(\pi) + (\delta-1)(1-\Upsilon(\pi)) \right] = -{\Psi} (\pi) + \lambda \left[ \Upsilon(\pi) - (1-\delta) \right] = L(\pi, \lambda)
\end{split}
\end{equation}
\center{\rule{18cm}{0.75pt}}
\end{table*}

\section{Proof of Lemma \ref{Lemma:OptimalityCase2}}\label{Appendix_Lemma:OptimalityCase2}
For the case that $\Upsilon(\pi_{\lambda})> 1-\delta$ when $\lambda=0$, we have $\lambda^*=0$ since $f'(\lambda)=\Upsilon({\pi}_{\lambda})  - (1-\delta)>0$ for all $\lambda\ge 0$.
For the case that $\Upsilon(\pi_{\lambda})< 1-\delta$ when $\lambda=0$, according to Lemma \ref{Lemma:ConvexEnvelope1} and Lemma \ref{Lemma:ConvexEnvelope2}, there must exist $j\in\{1,2,\cdots, Y-1\}$  such that $\Upsilon(\pi_{\lambda}) > 1-\delta$ if $\lambda \ge \lambda^{(j)}$, and $\Upsilon(\pi_{\lambda}) < 1-\delta$ otherwise.
Then, $\lambda^* = \lambda^{(j)}$ since $f'(\lambda)<0$ if $\lambda < \lambda^{(j)}$ and $f'(\lambda)>0$ otherwise.
Clearly, there is no other $\lambda'$ making $\Upsilon(\pi_{\lambda'}) > 1-\delta$ and $\Psi(\pi_{\lambda'})<\Psi(\pi_{\lambda^*})$ for both of the two cases.
\section{Proof of Theorem \ref{Theorem:RLPerformance}}\label{Appendix_Theorem:RLPerformance}
Under the reward design in \eqref{eq:RewardDesign}, we apply policy $\pi$ for  $K$ independent episodes, where stationary environment is considered such that $p(s',r|s,a)$ keep unchanged for the $K$ episodes for all $s'\in \mathcal{S}$, $r\in \mathcal{R}$, $s\in \mathcal{S}^{-}$,  and $a\in \mathcal{A}$.
Let $R_{t,k}(\pi)$ and $A_{t,k}(\pi)$ denote the reward received and action taken at time $t$ during the $k$th episode under policy $\pi$, respectively.
Similarly, $c_{t,k}(\pi)$ is the basic reward for successful data transmission  at time $t$ during the $k$th episode under policy $\pi$.
In addition, we define $I_k(\pi)$ to be the index of successful data transmission, that is $I_k(\pi)=1$ if the data transmission is successful in the $k$th episode under policy $\pi$ and $I_k(\pi)=0$ otherwise.

Under stationary environment,  the expected return  under a policy is the same as the average of return samples derived from an infinite number of tests under that policy.
Leveraging this fact, we can figure out the state value of the initial state $s_0$ under policy $\pi$ as
\setcounter{equation}{45}
\begin{equation}\label{eq:vPiS0eqLPiLambda_conclusion}
v_{\pi}(s_0) = L(\pi, \lambda),
\end{equation}
where the detailed deduction can be found in \eqref{eq:vPiS0eqLPiLambda_deduction} on the top of the next page.
It can be observed that the family of policies derived by the optimization of maximizing $L(\pi, \lambda)$ is the same with the family of policies derived by the reinforcement learning of maximizing $v_{\pi}(s_0)$. Recalling the definition of  $\pi_{\lambda}^{\dag}$ and $\pi_{\lambda}$, we have $\pi_{\lambda}^{\dag} = \pi_{\lambda}$.

\bibliographystyle{IEEEtran}
\bibliography{IEEEabrv,Reference}

\begin{thebibliography}{10}
\providecommand{\url}[1]{#1}
\csname url@samestyle\endcsname
\providecommand{\newblock}{\relax}
\providecommand{\bibinfo}[2]{#2}
\providecommand{\BIBentrySTDinterwordspacing}{\spaceskip=0pt\relax}
\providecommand{\BIBentryALTinterwordstretchfactor}{4}
\providecommand{\BIBentryALTinterwordspacing}{\spaceskip=\fontdimen2\font plus
\BIBentryALTinterwordstretchfactor\fontdimen3\font minus
  \fontdimen4\font\relax}
\providecommand{\BIBforeignlanguage}[2]{{%
\expandafter\ifx\csname l@#1\endcsname\relax
\typeout{** WARNING: IEEEtran.bst: No hyphenation pattern has been}%
\typeout{** loaded for the language `#1'. Using the pattern for}%
\typeout{** the default language instead.}%
\else
\language=\csname l@#1\endcsname
\fi
#2}}
\providecommand{\BIBdecl}{\relax}
\BIBdecl

\bibitem{2017-MVT-CoopITSinEurope}
K.~Sj$\rm{\ddot{o}}$berg, P.~Andres, T.~Buburuzan, and A.~Brakemeier,
  ``Cooperative intelligent transport systems in europe: Current deployment
  status and outlook,'' \emph{IEEE Veh. Technol. Mag.}, vol.~12, no.~2, pp.
  89--97, Jun. 2017.

\bibitem{2022-Network-AoILatencyReliability}
C.~Guo, X.~Wang, L.~Liang, and G.~Y. Li, ``Age of information, latency, and
  reliability in intelligent vehicular networks,'' \emph{IEEE Netw.}, {to be
  published, 2022}.

\bibitem{2012-CST-SurveyMACProtocolMissionCriticalApp}
P.~Suriyachai, U.~Roedig, and A.~Scott, ``A survey of {MAC} protocols for
  mission-critical applications in wireless sensor networks,'' \emph{IEEE
  Commun. Surveys Tuts.}, vol.~14, no.~2, pp. 240--264, Second Quarter 2012.

\bibitem{2018-Proc-URLLWirelessCommunTailRiskScale}
M.~Bennis, M.~Debbah, and H.~V. Poor, ``Ultrareliable and low-latency wireless
  communication: Tail, risk, and scale,'' \emph{IEEE Proc.}, vol. 106, no.~10,
  pp. 1834--1853, Oct. 2018.

\bibitem{2017-3GPP-22261}
3GPP, \emph{Service Requirements for the {5G} System}, document TS 22.261
  v16.0.0, 3rd Generation Partnership Project, June 2017.

\bibitem{2019-TCOM-JointPowerControlInSIMO}
O.~L.~A. L$\rm{\acute{o}}$pez, H.~Alves, and M.~Latva-aho, ``Joint power
  control and rate allocation enabling ultra-reliability and energy efficiency
  in {SIMO} wireless networks,'' \emph{IEEE Trans. Commun.}, vol.~67, no.~8,
  pp. 5768--5782, Aug. 2019.

\bibitem{2019-TCOM-WirelessAccessURLLC}
P.~Popovski, C.~Stefanovi$\rm{\acute{c}}$, J.~J. Nielsen, E.~Carvalho,
  M.~Angjelichinoski, K.~F. Trillingsgaard, and A.-S. Bana, ``Wireless access
  in ultra-reliable low-latency communication ({URLLC}),'' \emph{IEEE Trans.
  Commun.}, vol.~67, no.~8, pp. 5783--5801, Aug. 2019.

\bibitem{2018-JSAC-EnergyLatencyTradeoffInURLLC}
A.~Avranas, M.~Kountouris, and P.~Ciblat, ``Energy-latency tradeoff in
  ultra-reliable low-latency communication with retransmissions,'' \emph{IEEE
  J. Sel. Areas Commun.}, vol.~36, no.~11, pp. 2475--2485, Nov. 2018.

\bibitem{2021-WCL-LatSensitiveUAV}
S.~R. Pandey, K.~Kim, M.~Alsenwi, Y.~K. Tun, Z.~Han, and C.~S. Hong,
  ``Latency-sensitive service delivery with {UAV}-assisted {5G} networks,''
  \emph{IEEE Wireless Commun. Lett.}, vol.~10, no.~7, pp. 1518--1522, July
  2021.

\bibitem{2020-CL-MinmaxDecodingURLLC}
A.~A. Nasir, ``Min-max decoding-error probability-based resource allocation for
  a {URLLC} system,'' \emph{IEEE Commun. Lett.}, vol.~24, no.~12, pp.
  2864--2867, Dec. 2020.

\bibitem{2022-TCOM-URLLCedgeNet}
D.~V. Huynh, V.-D. Nguyen, S.~R. Khosravirad, V.~Sharma, O.~A. Dobre, H.~Shin,
  and T.~Q. Duong, ``{URLLC} edge networks with joint optimal user association,
  task offloading and resource allocation: A digital twin approach,''
  \emph{IEEE Trans. Commun.}, vol.~70, no.~11, pp. 7669--7682, Nov. 2022.

\bibitem{2023-TCOM-RAforCellFreeMIMO}
Q.~Peng, H.~Ren, C.~Pan, N.~Liu, and M.~Elkashlan, ``Resource allocation for
  uplink cell-free massive {MIMO enabled URLLC} in a smart factory,''
  \emph{IEEE Trans. Commun.}, vol.~71, no.~1, pp. 553--568, Jan. 2023.

\bibitem{2023-TCOM-PAinCellfreeUAV}
M.~Elwekeil, A.~Zappone, and S.~Buzzi, ``Power control in cell-free massive
  {MIMO} networks for {UAVs URLLC} under the finite blocklength regime,''
  \emph{IEEE Trans. Commun.}, vol.~71, no.~2, pp. 1126--1140, Feb. 2023.

\bibitem{2020-IoTJ-DRLbasedModeSelForV2X}
X.~Zhang, M.~Peng, S.~Yan, and Y.~Sun, ``Deep-reinforcement-learning-based mode
  selection and resource allocation for cellular {V2X} communications,''
  \emph{IEEE Internet Things J.}, vol.~7, no.~7, pp. 6380--6391, Jul. 2020.

\bibitem{2019-WCL-RAforV2XLargeDeviation}
C.~Guo, L.~Liang, and G.~Y. Li, ``Resource allocation for {V2X} communications:
  {A} large deviation theory perspective,'' \emph{IEEE Wireless Commun. Lett.},
  vol.~8, no.~4, pp. 1108--1111, Aug. 2019.

\bibitem{2019-TVT-DRLbasedRAforV2V}
H.~Ye, G.~Y. Li, and B.-H.~F. Juang, ``Deep reinforcement learning based
  resource allocation for {V2V} communications,'' \emph{IEEE Trans. Veh.
  Technol.}, vol.~68, no.~4, pp. 3163--3173, Apr. 2019.

\bibitem{2019-JSAC-SpectrumSharingMARL}
L.~Liang, H.~Ye, and G.~Y. Li, ``Spectrum sharing in vehicular networks based
  on multi-agent reinforcement learning,'' \emph{IEEE J. Sel. Areas Commun.},
  vol.~37, no.~10, pp. 2282--2292, Oct. 2019.

\bibitem{2020-Proc-DLwirelessRAwithVNET}
L.~Liang, H.~Ye, G.~Yu, and G.~Y. Li, ``Deep-learning-based wireless resource
  allocation with application to vehicular networks,'' \emph{IEEE Proc.}, vol.
  108, no.~2, pp. 341--356, Feb. 2020.

\bibitem{2019-TWC-RAforVehLowLatHighReliability}
C.~Guo, L.~Liang, and G.~Y. Li, ``Resource allocation for vehicular
  communications with low latency and high reliability,'' \emph{IEEE Trans.
  Wireless Commun.}, vol.~18, no.~8, pp. 3887--3902, Aug. 2019.

\bibitem{2023-TII-URLLCindIoT}
S.~Kurma, P.~K. Sharma, K.~Singh, S.~Mumtaz, and C.-P. Li, ``{URLLC-Based
  Cooperative Industrial IoT Networks with Nonlinear Energy Harvesting},''
  \emph{IEEE Trans. Ind. Inform.}, vol.~19, no.~2, pp. 2078--2088, Feb. 2023.

\bibitem{2022-ICC-AnalysisMEC}
S.~Suman, {\v{C}}.~Stefanovi\'{c}, S.~Do\v{s}en, and P.~Popovski, ``Analysis
  and optimization of the latency budget in wireless systems with mobile edge
  computing,'' in \emph{Proc. IEEE ICC}, Seoul, Korea, May 2022, pp.
  5029--5034.

\bibitem{2022-TCOM-TwoTimescale}
G.~Ding, J.~Yuan, G.~Yu, and Y.~Jiang, ``Two-timescale resource management for
  ultrareliable and low-latency vehicular communications,'' \emph{IEEE Trans.
  Commun.}, vol.~70, no.~5, pp. 3282--3294, May 2022.

\bibitem{2018-Book-ReinforcementLearning}
R.~S. Sutton and A.~G. Barto, \emph{Reinforcement Learning: An Introduction},
  2nd~ed.\hskip 1em plus 0.5em minus 0.4em\relax Cambridge, MA, USA: MIT Press,
  2018.

\bibitem{2004-Book-ConvexOpt}
S.~Boyd and L.~Vandenberghe, \emph{Convex Optimization}.\hskip 1em plus 0.5em
  minus 0.4em\relax Cambridge University Press, 2004.

\bibitem{2001-ConvergenceQlearning}
F.~S. Melo, ``Convergence of {Q}-learning: A simple proof,'' {Inst.
  Syst.Robot., Lisbon, Portugal, Tech. Rep., 2001, pp. 1-4}.

\bibitem{2019-TVT-MyTVT2019}
C.~Guo, L.~Liang, and G.~Y. Li, ``Resource allocation for high-reliability
  low-latency vehicular communications with packet retransmission,'' \emph{IEEE
  Trans. Veh. Technol.}, vol.~68, no.~7, pp. 6219--6230, July 2019.

\end{thebibliography}

\begin{IEEEbiography}[{\includegraphics[width=1in,height=1.25in,clip,keepaspectratio]{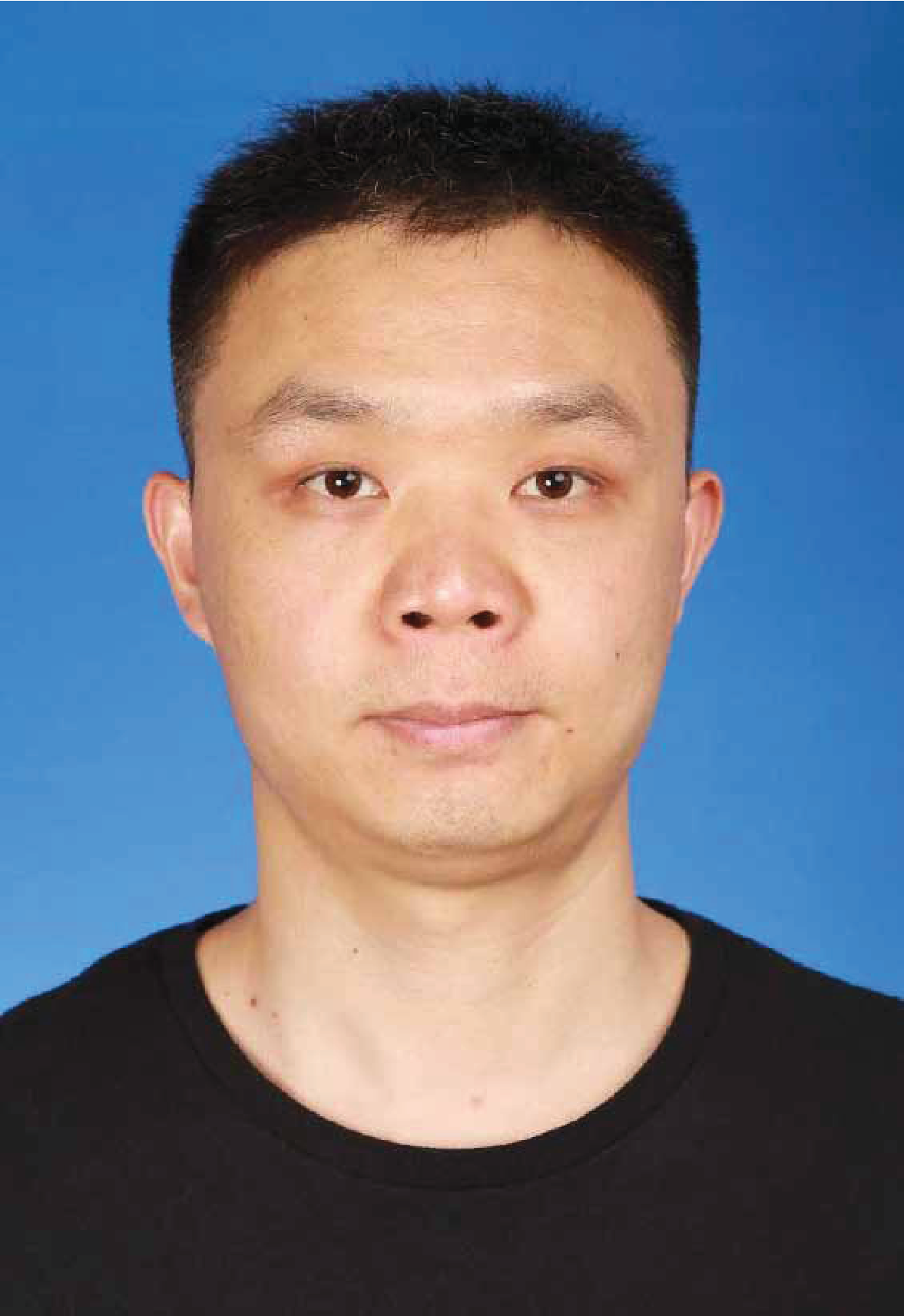}}]{Chongtao Guo} (M'14) received his B.Eng. degree and Ph.D. degree from Xidian University, Xi'an, China, in 2009 and 2014, respectively.

From 2017 to 2018, he was a post-doctoral research fellow   with Georgia Institute of Technology, Atlanta, GA, USA. Since 2014, he has been with Shenzhen University, Shenzhen, China, where he is currently an Associate Professor. His current research interests are in optimization and learning for wireless communications, vehicular networks, and Internet of Things.

Dr. Guo is a co-recipient of the Best Paper Award at 2016 21st International Conference on Digital Signal Processing and 2017 22nd International Conference on Digital Signal Processing.
\end{IEEEbiography}

\begin{IEEEbiography}[{\includegraphics[width=1in,height=1.25in,clip,keepaspectratio]{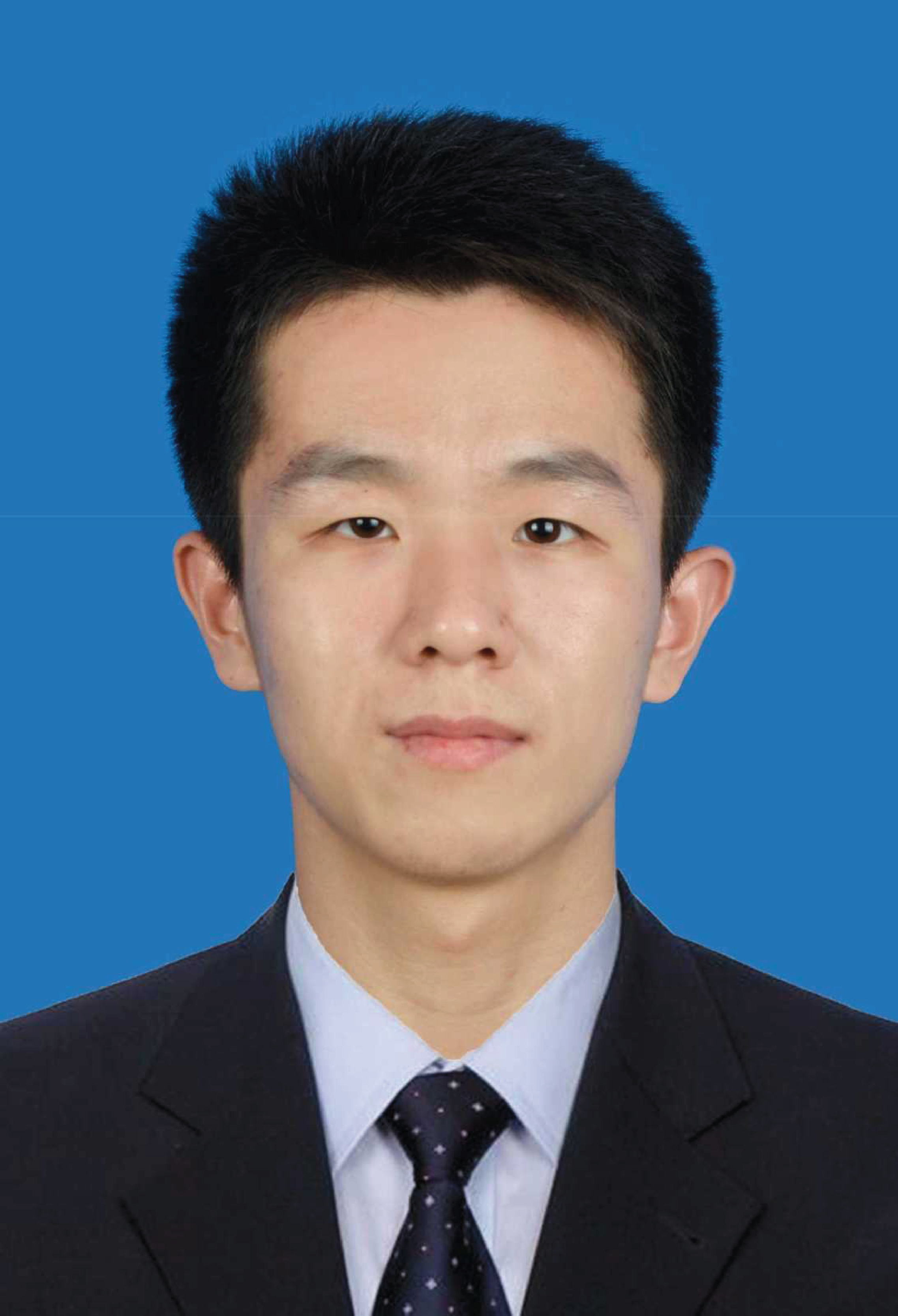}}]{Zhengchao Li} received the B.E. degree in electronic and information engineering from Nanyang Institute of Technology, Nanyang, China, in 2019 and the M.E. degree in electronic and communication engineering from Shenzhen University, Shenzhen, China, in 2022.
His main research interests are in wireless communications.

He currently works as a software development engineer in Huawei Technology Co., Ltd, Dongguan 523820, China.
\end{IEEEbiography}

\begin{IEEEbiography}[{\includegraphics[width=1in,height=1.25in,clip,keepaspectratio]{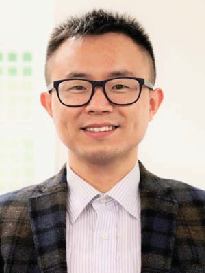}}]{Le Liang} (S'13-M'19) received the B.E. degree in information engineering from Southeast University, Nanjing, China, in 2012, the M.A.Sc degree in electrical engineering from the University of Victoria, Victoria, BC, Canada, in 2015, and the Ph.D. degree in electrical and computer engineering from the Georgia Institute of Technology, Atlanta, GA, in 2018. From 2019 to 2021, he was a Research Scientist at Intel Labs, Hillsboro, OR. Since 2021, he has been with the National Mobile Communications Research Laboratory, Southeast University, Nanjing, China. His main research interests are in wireless communications, signal processing, and machine learning.

Dr. Liang serves as an Associate Editor for the IEEE Transactions on Cognitive Communications and Networking and an Editor for the IEEE Communications Letters. He is a member of the Machine Learning for Signal Processing Technical Committee of the IEEE Signal Processing Society. He received the Best Paper Award of IEEE/CIC ICCC in 2014 and was named an Exemplary Reviewer of the IEEE Wireless Communications Letters in 2018.
\end{IEEEbiography}

\begin{IEEEbiography}[{\includegraphics[width=1in,height=1.25in,clip,keepaspectratio]{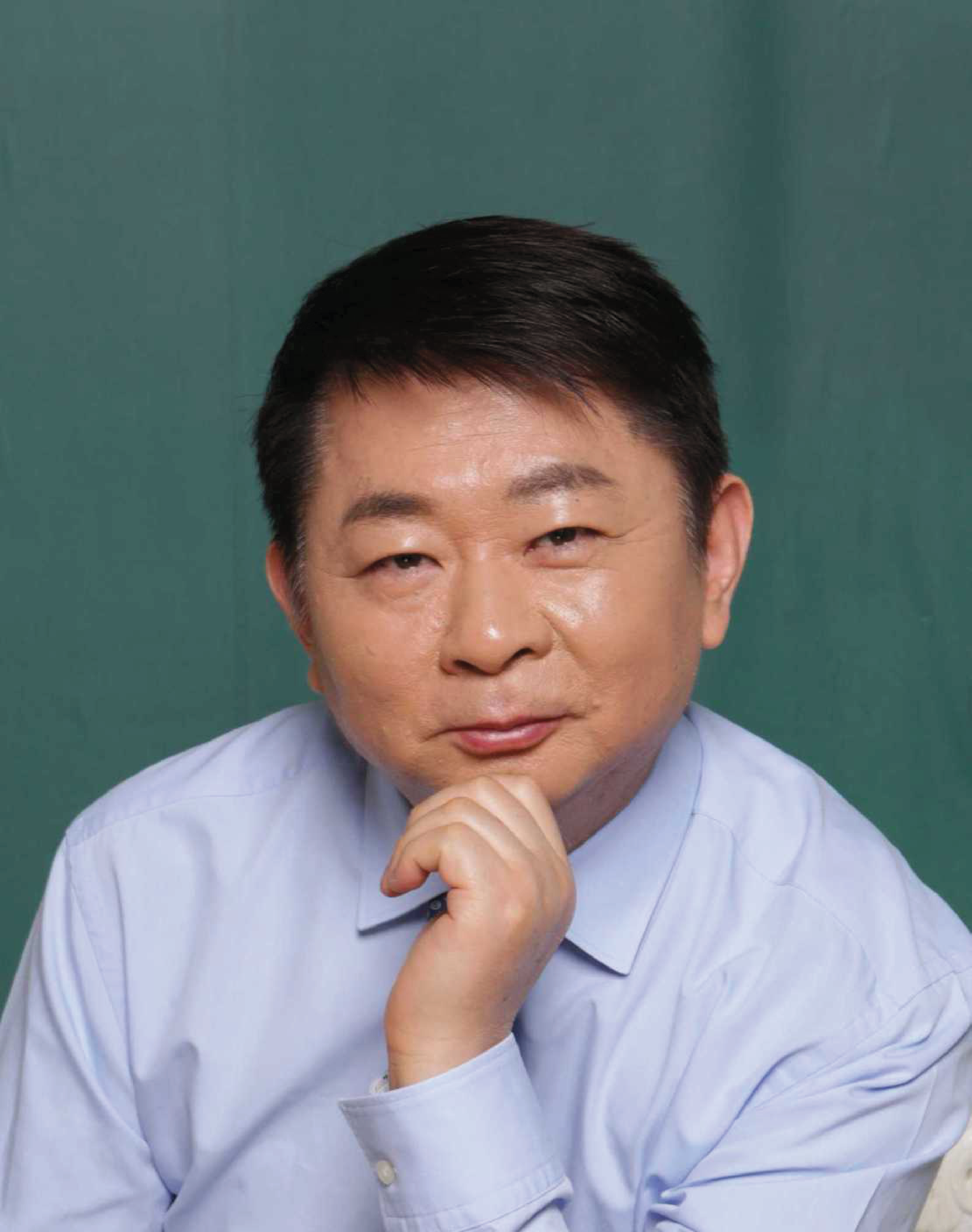}}]{Geoffrey Ye Li} is currently a Chair Professor at Imperial College London, UK.  Before joining Imperial in 2020, he was a Professor at Georgia Institute of Technology, USA, for 20 years and a Principal Technical Staff Member with AT\&T Labs - Research in New Jersey, USA, for five years. His general research interests include statistical signal processing and machine learning for wireless communications. In the related areas, he has published over 600 journal and conference papers in addition to over 40 granted patents and several books. His publications have been cited over 61,000 times with an H-index over 114 and he has been recognized as a Highly Cited Researcher, by Thomson Reuters, almost every year.

Dr. Geoffrey Ye Li was awarded IEEE Fellow and IET Fellow for his contributions to signal processing for wireless communications. He won several prestigious awards from IEEE Signal Processing, Vehicular Technology, and Communications Societies, including IEEE ComSoc Edwin Howard Armstrong Achievement Award in 2019.

He has been involved in editorial activities for over 20 technical journals, including the founding Editor-in-Chief of IEEE JSAC Special Series on ML in Communications and Networking. He has organized and chaired many international conferences, including technical program vice-chair of the IEEE ICC'03, general co-chair of the IEEE GlobalSIP'14, the IEEE VTC'19 Fall, the IEEE SPAWC'20, and the IEEE VTC'22 Fall.
\end{IEEEbiography}

\end{document}